\documentclass[aps,preprintnumbers,amsmath,amssymb,eqsecnum,floatfix,twocolumn]{revtex4}

\usepackage{graphicx}
\usepackage{euscript}
\usepackage{oldgerm}
\usepackage{subfigure}

\begin{document}

\title{Nanotube field of C$_\text{60}$ molecules in carbon nanotubes: atomistic versus continuous tube
approach}
\date{\today}
\author{B. Verberck and K.H. Michel}
\affiliation{Department of Physics, University of Antwerp, Groenenborgerlaan 171, 2020 Antwerpen, Belgium}

\begin{abstract}
We calculate the van der Waals energy of a C$_\text{60}$ molecule when it is encapsulated in a single-walled
carbon nanotube with discrete atomistic structure.  Orientational degrees of freedom and longitudinal
displacements of the molecule are taken into account, and several achiral and chiral carbon nanotubes are
considered.  A comparison with earlier work where the tube was approximated by a continuous cylindrical
distribution of carbon atoms is made.  We find that such an approximation is valid for high and intermediate tube
radii; for low tube radii, minor chirality effects come into play.  Three molecular orientational regimes are
found when varying the nanotube radius.
\end{abstract}

\maketitle

\section{Introduction}
The discovery of carbon nanotubes (CNTs) by Iijima \cite{Iij} and their subsequent large-scale production
\cite{Ebb} was followed by the synthesis of CNTs filled with atoms and/or molecules.
These novel hybrid materials
often exhibit one-dimensional characteristics and are presently the subject of fundamental studies as well as
research aiming at their application in nanotechnology.  For a review on CNTs and their filling we refer to
Refs.\ \onlinecite{Saibook,Har} and \onlinecite{Slo,Mon}, respectively.  Self-assembled chains of C$_\text{60}$
fullerene molecules inside single-walled carbon nanotubes (SWCNTs), the so-called peapods \cite{Smi}, provide a
unique example of such nanoscopic compound materials, and feature unusual electronic \cite{Hor} and structural
properties.  High-resolution transmission electron microscopy observations on CNTs filled sparsely with
C$_\text{60}$ molecules \cite{Smi2} demonstrate the motion of the fullerene molecules along the tube axis and
imply that the interaction between C$_\text{60}$ molecules and the surrounding nanotube wall is due to weak van
der Waals forces and not to chemical bonds.

Recently, the way the C$_\text{60}$ molecules of a (C$_\text{60}$)$_N$@SWCNT peapod \cite{Smi,Bur} --- $N$
C$_\text{60}$ molecules inside in a SWCNT --- are packed in the encapsulating tube has been investigated both
experimentally and theoretically \cite{Pickett,Hod1,Khl,Troche,MicVerNikPRL,MicVerNikEPJB}.  Obviously, the
structure of a peapod is governed by the interactions between
the C$_\text{60}$ molecules, and by the way a C$_\text{60}$ molecule interacts with the surrounding tube wall.
Already when considering the stacking of cylindrically confined hard spheres, a possible rudimentary description
of a (C$_\text{60}$)$_N$@SWCNT peapod, various chiral structures of the spheres stacking
for varying tube radius are obtained
\cite{Pickett}.  In Ref.\ \onlinecite{Hod1}, Hodak and Girifalco calculated lowest-energy
(C$_\text{60}$)$_N$@SWCNT peapod configurations by means of a continuum approach for the C$_\text{60}$-tube
interaction: both a SWCNT and a C$_\text{60}$ molecule are approximated as a homogeneous surface --- cylindrical
and spherical, respectively.  Although in doing so any effect of tube chirality and/or molecular
orientation can not be  accounted for, such a model provides useful information about the spatial arrangement of the spherical molecules in
the tube.  Ten different stacking arrangements were obtained for the tube radius
$R_\text{T}$ ranging from $6.27$ {\AA} to $13.57$ {\AA}.  The simplest configuration (C$_\text{60}$ ``spheres"
aligned linearly along the tube axis) occurs for the smallest tubes ($6.27$ {\AA} $\le R_\text{T}\le 7.25$ {\AA}).
Other phases consist of zig-zag patterns or C$_\text{60}$ balls forming helices.  Some of the predicted phases
have been observed experimentally \cite{Khl}.  Interestingly, experimental observations of similar structures
formed by C$_\text{60}$ molecules inside BN nanotubes have been reported as well \cite{Mickelson}.  An atomistic
molecular dynamics study on the arranging of C$_\text{60}$ molecules inside SWCNTs was carried out by Troche et
al.\ \cite{Troche}; the C$_\text{60}$-tube interaction was modelled by adding carbon-carbon Lennard-Jones 6-12
potentials.  Troche et al.\ \cite{Troche} concluded that the chirality of the encapsulating SWCNT has only a minor
effect on the lowest-energy configuration of the C$_\text{60}$ molecules and their obtained arrangements, thus
depending on the tube radius only, are in full agreement with those of Hodak and Girifalco \cite{Hod1}.
Conclusions on the individual orientations of C$_\text{60}$ molecules inside a SWCNT were not given by Troche et
al.\ \cite{Troche} --- their goal was to study the packing of several molecules.  Molecular orientation effects
are expected to come into play at sufficiently low temperatures when orientational motion is frozen, and indeed do
so as was shown in Refs.\ \onlinecite{MicVerNikPRL} and \onlinecite{MicVerNikEPJB}, where the potential energy
of a single C$_\text{60}$ molecule confined to the tube axis of a SWCNT, called ``nanotube field", was calculated
by treating the tube as a homogeneous cylindrical carbonic surface density but retaining the icosahedral features
of a C$_\text{60}$ molecule.  A specific dependence on the tube radius was found; three distinct molecular
orientations were observed within the range $6.5\lesssim R_\text{T}\lesssim 8.5$ {\AA}.  It is our opinion that,
for calculating tube-C$_\text{60}$ interactions, taking the detailed molecular structure of a C$_\text{60}$
molecule into account has priority over the chiral structure of a nanotube.  Replacing a SWCNT by a continuous
cylindrical distribution of carbon atoms is intuitively justifiable, but treating a C$_\text{60}$ molecule as a
sphere (as in Ref.\ \cite{Hod1}) with no further structure is a more
questionable approximation.  Indeed, whereas the
carbon-carbon bonds in a CNT are of one type, a C$_\text{60}$ molecule features longer (``single") and shorter
(``double") bonds, arranged in pentagons --- electron-poor regions --- and hexagons --- electron-rich regions.
The importance of taking the detailed molecular structure properly into account follows from Refs.\
\onlinecite{MicVerNikPRL} and \onlinecite{MicVerNikEPJB}; but the neglect of the discrete atomistic structure of the tube when
considering C$_\text{60}$-tube interactions, although intuitively plausible, requires solid grounds.
The goal of this paper is to answer the question how good a smooth-tube approximation really is, and to confirm
the relevance of the precise structure of a
C$_\text{60}$ molecule, i.e.\ the importance of allowing for molecular orientational degrees of freedom.

The content of the paper is as follows.  In Sec.\ \ref{nanotubefield}, we discuss formulas for the calculation of
the nanotube field of an encapsulated C$_\text{60}$ molecule for both a ``continuous" and a ``discrete" tube.
Then (Sec.\ \ref{comparison}), we plot nanotube fields for a selection of
representative nanotubes and make preliminary visual comparisons between the two approaches.
In Sec.\ \ref{casestudy}, we present an all-variable treatment and apply it for tubes with intermediate and small
tube radii.  Finally, general conclusions are given (Sec.\ \ref{discussion}).

\section{Nanotube field}\label{nanotubefield}
\begin{figure}
\resizebox{!}{6cm}{\includegraphics{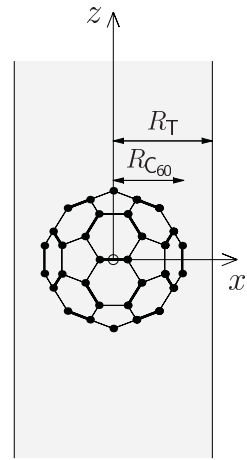}}
\caption{A single C$_\text{60}$ molecule (radius $R_\text{C$_\text{60}$}$) in a SWCNT (radius $R_\text{T}$).
Shown is a projection onto the $(x,z)$-plane.  The center of mass of the C$_\text{60}$ molecule is chosen as the
coordinate system's origin.  The tube's long axis coincides with the $z$-axis; the C$_\text{60}$ molecule is put
in the standard orientation.\label{figswcnt}}
\end{figure}

\begin{figure*}
\subfigure{\resizebox{5cm}{!}
{\includegraphics{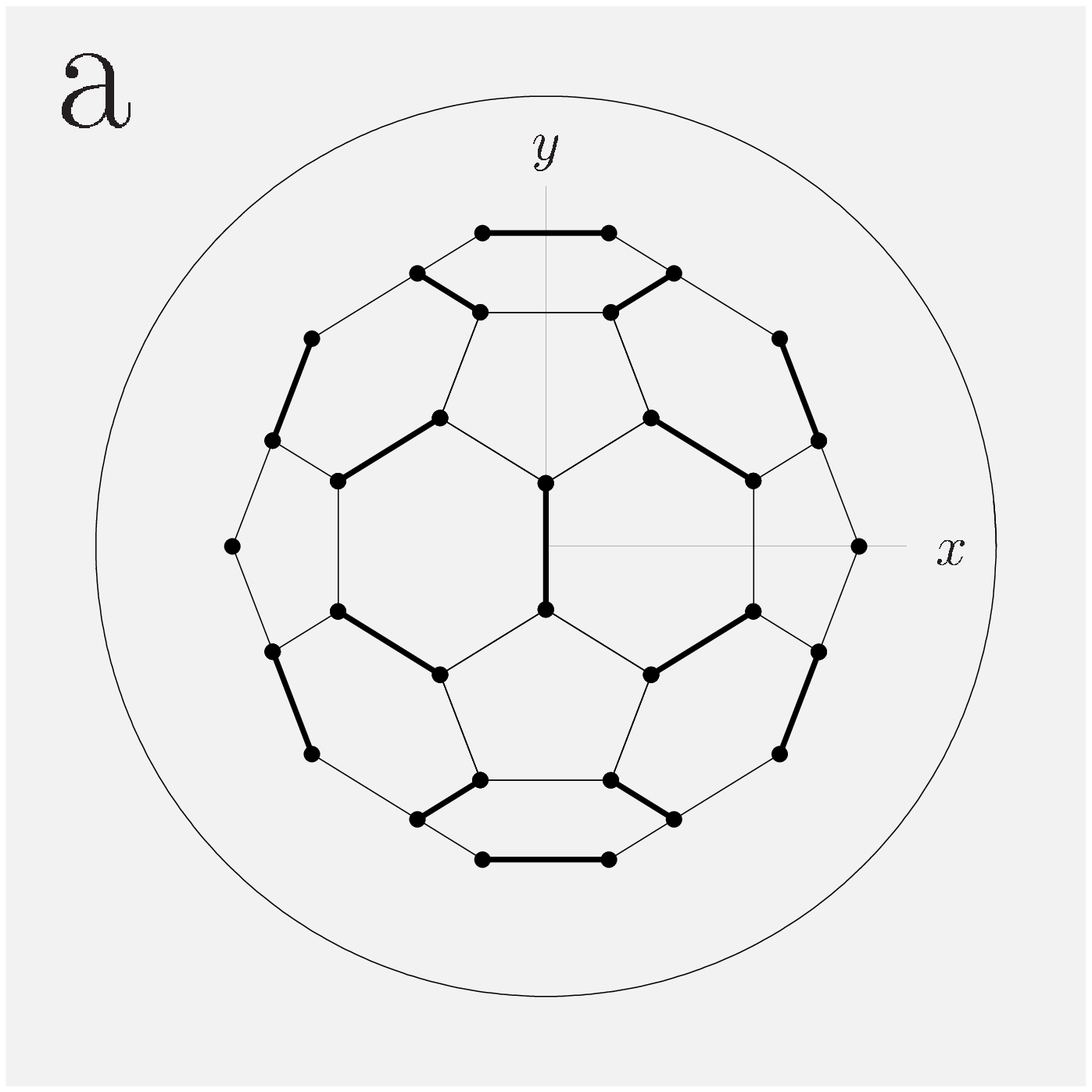}}}
\subfigure{\resizebox{5cm}{!}
{\includegraphics{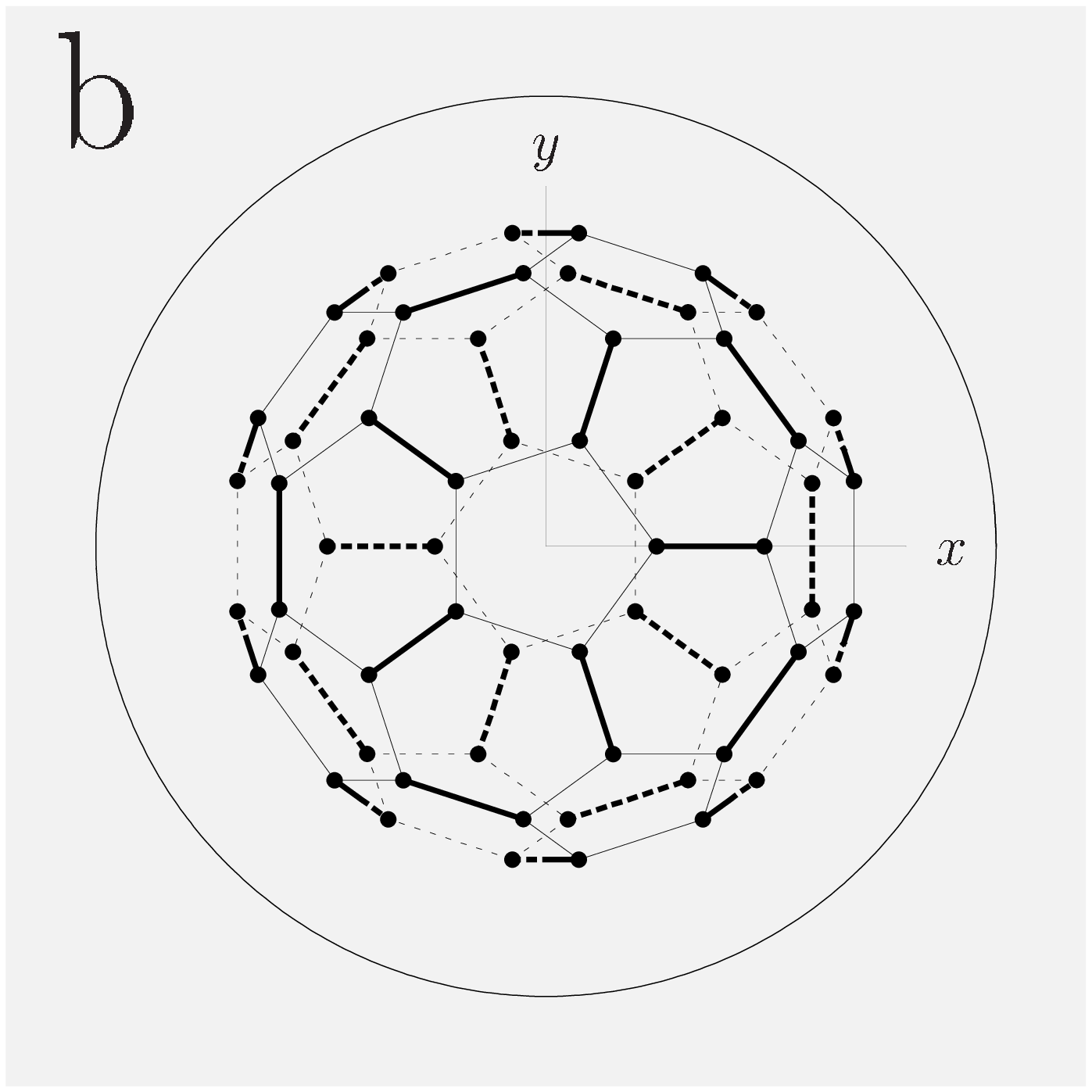}}}
\subfigure{\resizebox{5cm}{!}
{\includegraphics{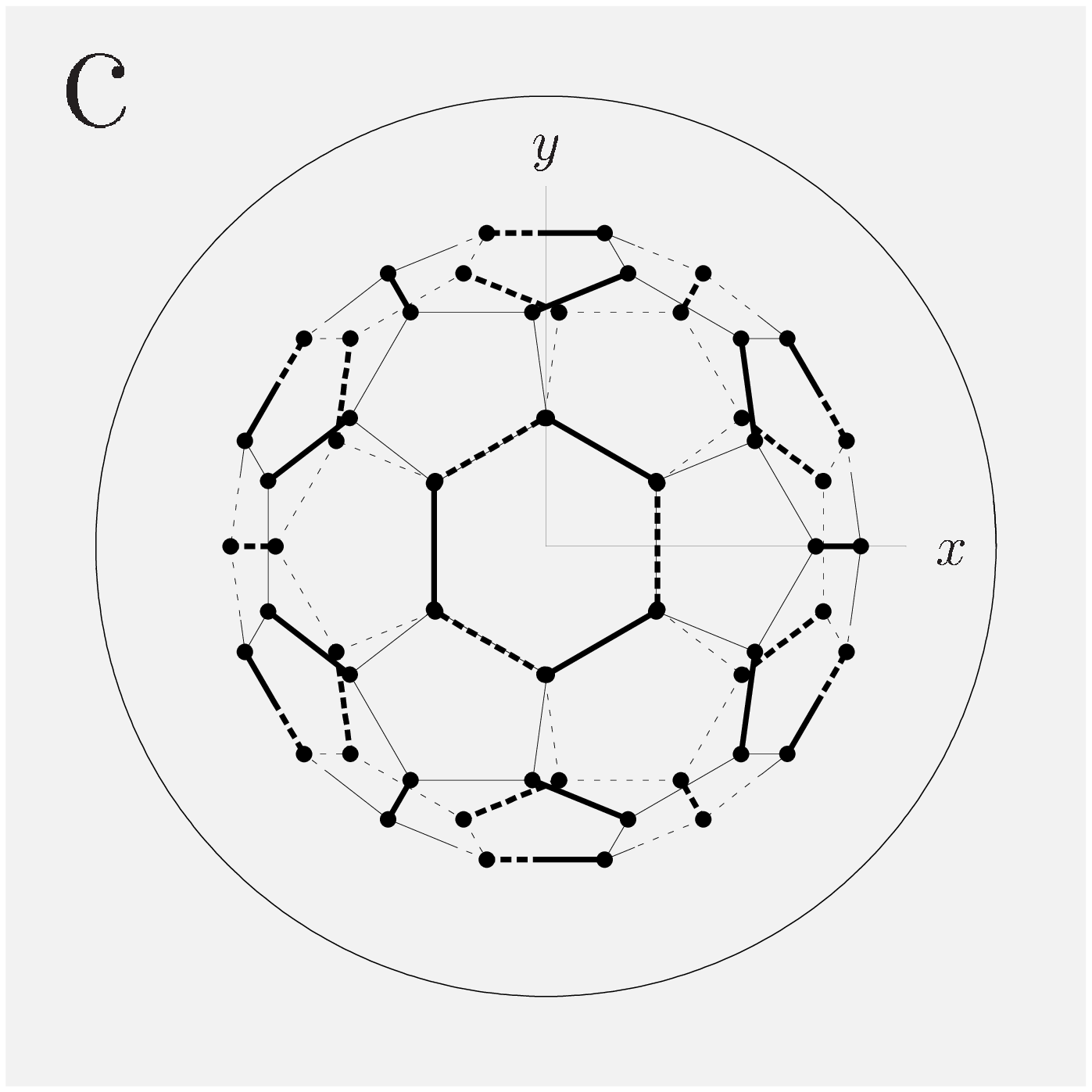}}}
\caption{
Special orientations of the C$_\text{60}$ molecule: (a) standard orientation ($\alpha=\beta=\gamma=0$, double
bonds perpendicular to the $z$-axis), (b) ``pentagonal" orientation (pentagons perpendicular to the $z$-axis), (c)
``hexagonal" orientations (hexagons perpendicular to the $z$-axis).  Double bonds are drawn bolder than single
bonds; dashed lines are located ``beneath" the $(x,y)$-plane ($z\le0$).  The surrounding tube is represented as a
circular projection onto the $(x,y)$-plane.  Note that a rotation about the $z$-axis over $-\alpha$ does not
change double bonds, pentagons or hexagons --- for Figs.\ (a), (b) and (c), respectively --- being perpendicular
to the $z$-axis.}
\label{orientations}
\end{figure*}

We consider a C$_\text{60}$ molecule in a SWCNT, the molecule assuming a centered position in the tube, and set up
a cartesian system of axes $(x,y,z)$ so that the $z$-axis coincides with the tube's long axis and contains the
molecule's center of mass (Fig.\ \ref{figswcnt}).  The potential energy $V$ of the C$_\text{60}$ molecule then
depends on the orientation of the molecule, which can be characterized by three Euler angles
$(\alpha,\beta,\gamma)$, on the position of the molecule along the tube, i.e.\ the $z$-coordinate of the molecular
center of mass for which we write $\zeta$, and on the tube indices \cite{Saibook} $(n,m)$:
\begin{align}
   V\equiv V(\alpha,\beta,\gamma;\zeta;n,m).
\end{align}
For the Euler angles we use the convention of Ref.\ \cite{BraCrack}: a coordinate function
$f\bigl(\vec{r}=(x,y,z)\bigr)$ is transformed as
${\mathfrak R}(\alpha,\beta,\gamma)f(\vec{r}\,)=f\bigl({\mathfrak R}^{-1}(\alpha,\beta,\gamma)\vec{r}\,\bigr)$,
where ${\mathfrak R}(\alpha,\beta,\gamma)={\mathfrak R}_z(\gamma){\mathfrak R}_y(\beta){\mathfrak R}_z(\alpha)$
stands for the succession of a rotation over $0\le\alpha<2\pi$ about the $z$-axis, a rotation over
$0\le\beta\le\pi$ about the $y$-axis, and a rotation over $0\le\gamma<2\pi$ about the $z$-axis again.  The $x$-,
$y$- and $z$-axes are kept fixed.  Note that the {\bf coordinate} transform associated with the Euler angles reads
$\vec{r}\,'={\mathfrak R}^{-1}(\alpha,\beta,\gamma)\vec{r}={\mathfrak R}_z(-\alpha){\mathfrak R}_y(-\beta)
{\mathfrak R}_z(-\gamma)\vec{r}$ and that the rotation of the C$_\text{60}$ molecule over $-\alpha$ about the
$z$-axis is performed last.  As the starting orientation [$(\alpha=0,\beta=0,\gamma=0)$] we take the so-called
standard orientation [Fig.\ \ref{orientations}(a)]: twofold molecular symmetry axes then coincide with the
cartesian axes and every cartesian axis intersects two opposing double bonds.  (We recall that the carbon-carbon
bonds of a C$_\text{60}$ molecule can be divided into two categories: 60 single bonds, fusing pentagons and
hexagons, and 30 double bonds, fusing hexagons.  The latter are somewhat longer than the former \cite{DavNature}.)
Bearing in mind the results of Refs.\ \onlinecite{MicVerNikPRL} and \onlinecite{MicVerNikEPJB} and anticipating
the results obtained in the present work, we point out two more molecular orientations of importance.  The first
is the ``pentagonal" orientation class, obtained by the Euler transformation
$(\alpha\text{ arbitrary},\beta=\cos^{-1}\frac{2}{\sqrt{10+2\sqrt{5}}}\approx 58^\circ,\gamma=0)$, resulting in
two opposing pentagons of the C$_\text{60}$ molecule being perpendicular to the $z$-axis [Fig.\
\ref{orientations}(b)].  The second is the category of ``hexagonal" orientations, a result of the Euler
transformation $(\alpha\text{ arbitrary},\beta=\cos^{-1}\frac{1+\sqrt{5}}{2\sqrt{3}}\approx 21^\circ,\gamma=0)$,
making two opposing hexagons lie perpendicular to the $z$-axis [Fig.\ \ref{orientations}(c)].  The angle
$\beta_0 = \cos^{-1}\frac{1+\sqrt{5}}{2\sqrt{3}}$ is related to the dihedral angle $\psi$ (the inner angle between
adjacent faces) of a regular icosahedron: $\psi = \pi - 2\beta_0$.  Other $(\beta,\gamma)$ pairs yield
``pentagonal", ``hexagonal" and ``double-bond" orientations as well: 12 pairs correspond to a ``pentagonal", 20
pairs to a ``hexagonal", and 30 pairs to a ``double-bond" orientation since a C$_\text{60}$ molecule has 12
pentagons, 20 hexagons and 30 double bonds.  

For the description of the interaction between the C$_\text{60}$ molecule and the nanotube we follow earlier work
\cite{LamZeitschrift} and treat the C$_\text{60}$ molecule as a rigid cluster of interaction centers (ICs).  Not
only C atoms (`a') act as ICs, but also double bonds (`db') and single bonds (`sb').  We label the 60 atoms by the
index $\lambda_\text{a}=1,\hdots,60$.  In the center of every of the 60 single bonds an IC is put, labelled by the
index $\lambda_\text{sb} = 1,\hdots,60$.  On each of the 30 double bonds, 3 ICs dividing the bond in four equal
parts are put, totalling to 90 db ICs, labelled $\lambda_\text{db} = 1,\hdots,90$.
Such a construction was originally introduced for modelling intermolecular interactions in solid C$_\text{60}$
(C$_\text{60}$ fullerite); having three ICs per double bond reflects the electronic density being smeared out
along a double bond \cite{LamZeitschrift}.

Every IC of the C$_\text{60}$
molecule interacts with every atom of the nanotube via a pair interaction potential
$v^\text{t}\bigl(|\vec{r}_\tau-\vec{r}_{\lambda_\text{t}}|\bigr)$, depending on the type of IC (`t' = `a', `db',
`sb').  The total potential energy is then obtained by summing over all pair interactions:
\begin{subequations}
\begin{align}
   V = \sum_\tau\sum_{\text{`t'}=\text{`a'},\text{`db'},\text{`sb'}}
                \sum_{\lambda_\text{t}}v^\text{t}\bigl(|\vec{r}_\tau-
\vec{r}_{\lambda_\text{t}}|\bigr),
		\label{discreteeq}
\end{align}
where $\tau$ indexes the atoms of the tube and $\vec{r}_\tau$ stands for their respective coordinates.  As in
Refs.\ \onlinecite{MicVerNikPRL} and \onlinecite{MicVerNikEPJB}, we use Born--Mayer--van der Waals pair
interaction potentials:
\begin{align}
   v^\text{t}(r) = C_1^te^{-C_2^tr}-\frac{B^t}{r^6}. \label{ppotential}
\end{align}
Again, the use of such pair potentials was originally introduced for studying C$_\text{60}$ - C$_\text{60}$
interactions in C$_\text{60}$ fullerite \cite{LamZeitschrift}; it lead to a crystal field potential and a
structural phase transition temperature \cite{124,130} in good agreement with experiments.
 The potential constants $C_1^t$, $C_2^t$ and $B^t$ used are those of Ref.\ \onlinecite{MicVerNikEPJB}.  In Eq.\
(\ref{discreteeq}), the sum over tube atoms, labelled by the index $\tau$ and having coordinates
$\vec{r}_\tau=(x_\tau,y_\tau,z_\tau)$, can be restricted to atoms in a certain vicinity of the C$_\text{60}$
molecule, realized by imposing the criterion
\begin{align}
   z_\text{min}\le z_\tau\le z_\text{max}, \label{cutoff}
\end{align}
\end{subequations}
with $z_\text{min}$ and $z_\text{max}$ cut-off values ensuring convergence.

In Refs.\ \cite{MicVerNikPRL} and \cite{MicVerNikEPJB}, a smooth-tube 
approximation to Eq.\ (\ref{discreteeq})
was presented.  The actual network of carbon atoms making up the SWCNT is 
replaced by a homogeneous, cylindrical
``carbonic" surface density with value $\sigma$ (units {\AA}$^{-2}$).  The 
C$_\text{60}$ molecule-nanotube
interaction energy is then rewritten as
\begin{multline}
   V_\text{smooth} \\ = \sigma R_\text{T}\int_0^{2\pi}d\Phi\int_{-
\infty}^{+\infty}dZ
\sum_{\text{`t'}=\text{`a'},\text{`db'},\text{`sb'}}\sum_{\lambda_\text{t}}v^
\text{t}
\bigl(|\vec{\rho}-\vec{r}_{\lambda_\text{t}}|\bigr), \label{smootheq}
\end{multline}
where $\vec{\rho}=(R_\text{T},\Phi,Z)$ is the cylindrical coordinate of a 
point on the tube
($x=R_\text{T}\cos\Phi$, $y=R_\text{T}\sin\Phi$, $z=Z$) and $R_\text{T}$ is 
the tube radius.  The motivation for
introducing approximation (\ref{smootheq}) is twofold.  One reason is the 
dependence of $V_\text{smooth}$ on
the tube radius $R_\text{T}$ rather than on the tube indices $(n,m)$.  
Indeed, $R_\text{T}$ remains the only
relevant tube-characteristic parameter and as such simplifies a systematic 
investigation of carbon nanotubes.  A
further consequence of the tube's cylindrical symmetry is the irrelevance of the 
Euler angle $\alpha$ (a final
rotation of the C$_\text{60}$ molecule over $-\alpha$ about the tube axis 
doesn't matter) and of the $z$-coordinate
$\zeta$ (for infinite or long-enough tubes).  A second advantage of the 
smooth-tube ansatz is the possibility of
performing an expansion of $V_\text{smooth}$ into symmetry-adapted rotator 
functions, a point we will
return to in Sec.\ \ref{discussion}.  We stress the limited dependence of 
$V_\text{smooth}$ by writing
\begin{align}
   V_\text{smooth}\equiv V_\text{smooth}(\beta,\gamma;R_\text{T}).
\end{align}
To distinguish the smooth-tube approximation from the discrete case, we add 
the subscript `discrete':
\begin{align}
   V\equiv V_\text{discrete}(\alpha,\beta,\gamma;\zeta;n,m),
\end{align}
where the actual expression is given by Eqs.\ (\ref{discreteeq}) -- 
(\ref{cutoff}).

In this paper we test the validity of smooth-tube approximation 
(\ref{smootheq}) by comparing $V_\text{discrete}$
and $V_\text{smooth}$ for a selection of tubes.  Bearing in mind the three 
qualitatively different radii ranges
($R_\text{T}\lesssim 7$ {\AA}, $7$ {\AA} $\lesssim R_\text{T}\lesssim 7.9$ 
{\AA} and $7.9$ {\AA}
$\lesssim R_\text{T}$) obtained in Ref.\ \cite{MicVerNikEPJB}, we have selected
zig-zag, armchair and chiral tubes with radii around $R_\text{T}=6.5$ {\AA}, 
$R_\text{T}=7.5$ {\AA} and
$R_\text{T}=8.5$ {\AA}.  We have generated $(n,m)$ tubes starting from a 
graphene sheet with basis vectors
$\vec{a}_1=a\vec{e}_X$ and $\vec{a}_2 = a\frac{1}{2}\vec{e}_X + 
a\frac{3}{2}\vec{e}_Y$, where $\vec{e}_X$ and
$\vec{e}_Y$ are planar cartesian basis vectors, and performing the roll-up 
along the vector
$\vec{C}(n,m)=n\vec{a}_1+m\vec{a}_2$ \cite{Ham,Saibook}.  The tube is then positioned 
so that the C atom originally (before rolling
up) at $0\vec{e}_X+0\vec{e}_Y$ lies in the $(x,y)$ plane with $x$-coordinate $0$ and $y$-coordinate 
$R_\text{T}$
and that the cylinder containing the C atoms has its long
axis coinciding with the $z$-axis.
The C$_\text{60}$ molecule is initially positioned so that its center of mass 
lies at the origin ($\zeta = 0$); a
translation along the $z$-axis away from the initial position is measured via 
the center of
mass' $z$-coordinate $\zeta$.  The radius of the tube with indices $(n,m)$
reads $R_\text{T}=\frac{a}{2\pi}\sqrt{n^2+nm+m^2}$, with $a = 
2.49$ {\AA} \cite{Ham,Saibook}; the
corresponding surface density has the value
\begin{align}
   \sigma = \frac{4}{\sqrt{3}a^2} = 0.372\text{ {\AA}$^{-2}$}. 
\label{sigmaeq}
\end{align}
A further tube parameter is its translational
periodicity $\Delta z$, relevant when considering the $\zeta$-dependence of
V$_\text{discrete}$.  While $\Delta z$ is small for non-chiral --- i.e.\ zig-zag, $\Delta z = \sqrt{3}a$, and
armchair, $\Delta z = a$ --- tubes, the translational period can get very large
for chiral tubes \cite{Saibook}.  A tube may also have an $s$-fold symmetry axis (coinciding with the
$z$-axis) and therefore a rotational period $\Delta\alpha = 2\pi/s$.  When considering a tube with 
$s$-fold rotational symmetry it suffices to examine the interval $0\le\alpha<\Delta\alpha$.
The periodicities and other tube characteristics of our selected tubes are listed
in Table \ref{tubes}.

\begin{table*}
\caption{
Characteristics of selected $(n,m)$ tubes.  Tubes of all types (zig-zag, chiral and armchair) with radii
$R_\text{T}$ as close to $6.5$ {\AA}, $7.5$ {\AA} and $8.5$ {\AA} as possible were chosen.
The angle $\Delta\alpha$ is the rotational period of the tube when performing a rotation
about the $z$-axis.
  An expression for the
calculation of the translational periodicity $\Delta z$ is given in Ref.\ \onlinecite{Saibook}.
}
\label{tubes}
\begin{ruledtabular}
\begin{tabular}{rrrrr}
$(n,m)$ & chirality & $R_\text{T}$ ({\AA})& $\Delta \alpha$ & $\Delta z$ ({\AA}) \\
\hline
$(16,0)$  & zig-zag  & $6.3407$ & $2\pi/16$ & $\sqrt{3}a = 4.3128$ \\
$(14,4)$  & chiral   & $6.4876$ & $2\pi/2$  & $35.3018$ \\
$(9,9)$   & armchair & $6.1176$ & $2\pi/9$  & $a = 2.49$ \\
\hline
$(19,0)$  & zig-zag  & $7.5296$ & $2\pi/19$ & $\sqrt{3}a$ \\
$(16,5)$  & chiral   & $7.5296$ & $2\pi$            & $81.9433$ \\
$(11,11)$ & armchair & $7.5504$ & $2\pi/11$ & $a$ \\
\hline
$(21,0)$  & zig-zag  & $8.3222$ & $2\pi/21$ & $\sqrt{3}a$ \\
$(21,1)$  & chiral   & $8.5273$ & $2\pi$            & $92.8005$ \\
$(12,12)$ & armchair & $8.2369$ & $2\pi/12$ & $a$ \\
\end{tabular}
\end{ruledtabular}
\end{table*}



\section{Mercator maps}\label{comparison}
To get a preliminary idea of how $V_\text{smooth}$ and $V_\text{discrete}$ compare, we have simply
plot $V_\text{smooth}(\beta,\gamma,R_\text{T})$ and $V_\text{discrete}(\alpha = 0,\beta,\gamma;\zeta = 0;n,m)$ for
each of the selected $(n,m)$ tubes in the form of Mercator maps \cite{Mercator}.  We stress that
$(\alpha = 0,\zeta = 0)$ is but a particular case and that final conclusions should be made not only on the
variation of $\beta$ and $\gamma$ but on the varying of $\alpha$ and $\zeta$ as well, as we will do later on.  We
do point out, however, that we expect the $\alpha$- and $\zeta$-dependencies to be of a lesser magnitude than the
$\beta$- and $\gamma$-dependencies since the former correspond to a (final) rotation of the molecule about the
$z$-axis (tube axis) over $-\alpha$ and a translation of the molecule along the $z$-axis, respectively, and hence
relate to the tube structure rather than to the molecule structure.  (As argued in the Introduction, a carbon
nanotube can be regarded as being more ``continuous" than a C$_\text{60}$ molecule.)

As for the cut-off values, we have found --- for $\zeta = 0$ --- that $z_\text{min} = -50$ {\AA} and
$z_\text{max} = 50$ {\AA} yield sufficient convergence.  Note that the choice of $z_\text{min}$ and $z_\text{max}$
fixes the numbers $N_\text{T}$ of atoms to be taken into account in sum (\ref{discreteeq}).  In principle, the
tube fragment of length $L=z_\text{max}-z_\text{min}$ has a surface density
$\tilde{\sigma}=N_\text{T}/(2\pi R_\text{T}L)$, differing from $\sigma$.  We observe that
differences between $\sigma$ and $\tilde{\sigma}$ are small, however.
Although possibly (slightly) enhancing the agreement between $V_\text{discrete}$ and $V_\text{smooth}$,
we have chosen not to calculate $V_\text{smooth}$ with
$\tilde{\sigma}$ since it somehow relates to the tube structure --- $N_\text{T}$
depends on $(n,m)$ ---
hence
surpassing the smooth-tube approach's underlying basic idea ($R_\text{T}$-dependence rather than
$(n,m)$-dependence).

\begin{figure*}
\resizebox{15cm}{!}{\includegraphics{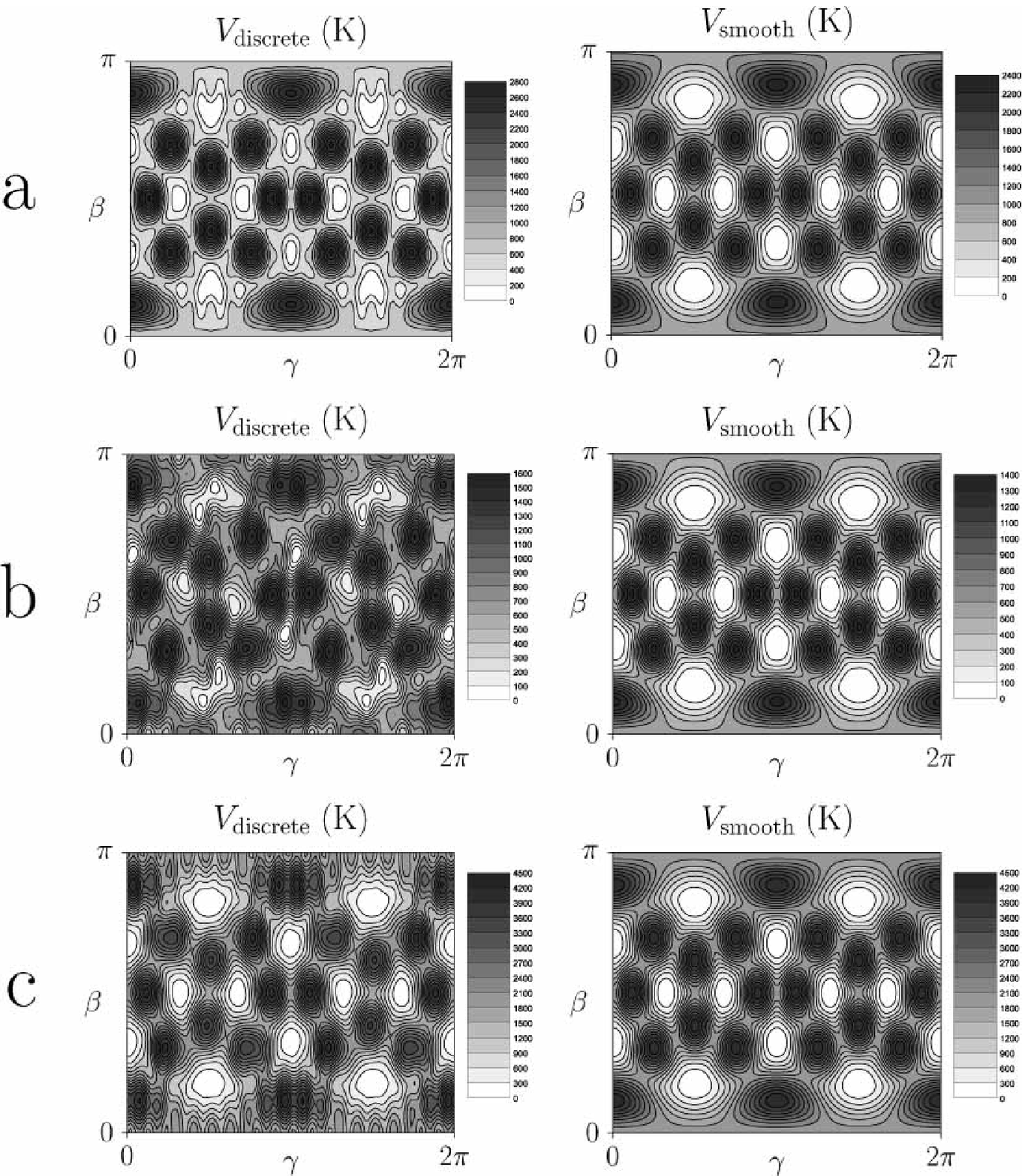}}
\caption{Mercator maps $V_\text{discrete}(\beta,\gamma)$ (left) and $V_\text{smooth}(\beta,\gamma)$ (right), units
K: (a) $(n,m) = (16,0)$, (b) $(n,m) = (14,4)$, (c) $(n,m) = (9,9)$.  The absolute minima values have been
subtracted.}
\label{figMercator1}
\end{figure*}

\begin{figure*}
\resizebox{15cm}{!}{\includegraphics{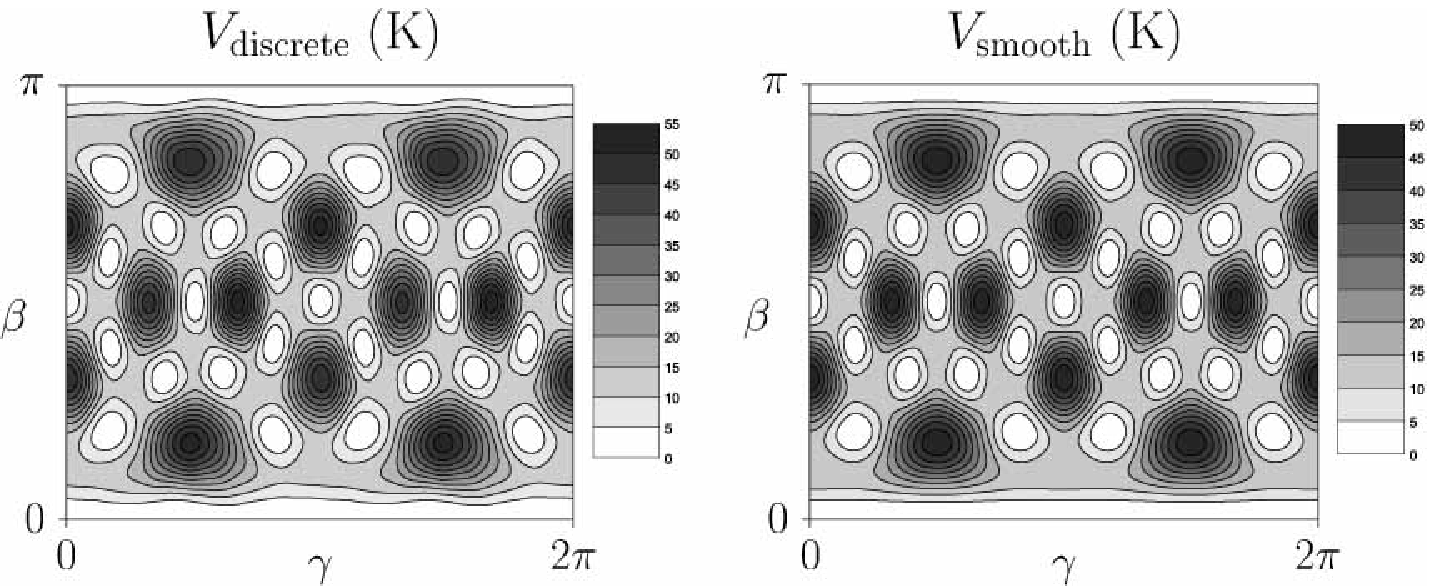}}
\caption{Mercator maps $V_\text{discrete}(\beta,\gamma)$ (left) and $V_\text{smooth}(\beta,\gamma)$ (right), units
K, for the $(n,m) = (16,5)$ tube.  The minimal values have been subtracted.  The maps for
the cases $(n,m) = (19,0)$ and $(11,11)$ look very similar to the $(16,5)$ maps.}
\label{figMercator2}
\end{figure*}

\begin{figure*}
\resizebox{15cm}{!}{\includegraphics{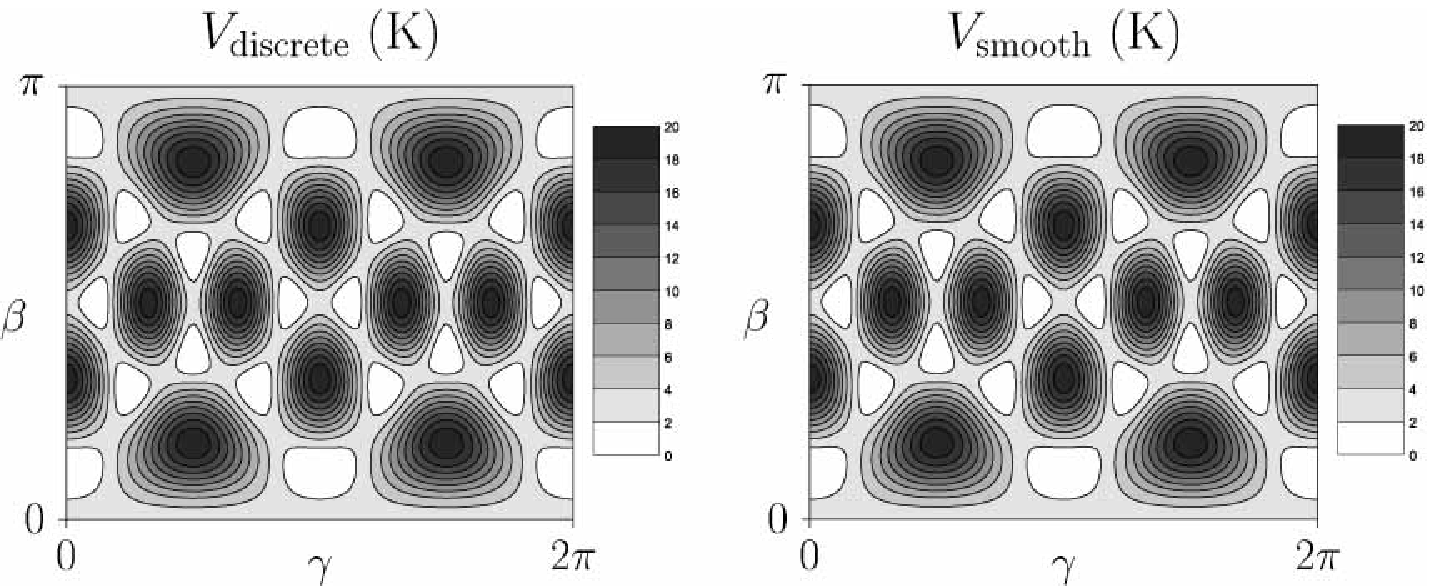}}
\caption{Mercator maps $V_\text{discrete}(\beta,\gamma)$ (left) and $V_\text{smooth}(\beta,\gamma)$ (right), units
K, for the $(n,m) = (21,1)$ tube.  The minimal values have been subtracted.  The maps for
the cases $(n,m) = (21,0)$ and $(12,12)$ are very similar to the $(21,1)$ maps.}
\label{figMercator3}
\end{figure*}

The Mercator maps $V_\text{smooth}(\beta,\gamma)\equiv V_\text{smooth}(\beta,\gamma;R_\text{T})$ and
$V_\text{discrete}(\beta,\gamma)\equiv V_\text{discrete}(\alpha=0,\beta,\gamma;\zeta = 0;n,m)$, respectively
calculated via Eqs.\ (\ref{discreteeq}) -- (\ref{cutoff}) and (\ref{smootheq}) and both based on pair potential
(\ref{ppotential}), are shown as Figs.\ \ref{figMercator1} -- \ref{figMercator3} for the tubes listed in Table
\ref{tubes}.  Figs.\ \ref{figMercator1}, \ref{figMercator2} and \ref{figMercator3} are for tubes with radii around
$6.5$ {\AA}, $7.5$ {\AA} and $8.5$ {\AA}, respectively.  Within each figure, Subfigs.\ (a), (b) and (c) refer to
zig-zag, chiral and armchair tubes; the left plot is $V_\text{discrete}$, the right $V_\text{smooth}$.  Only the
variation is plotted; for each plot the lowest occurring energy value, for which we write $V^0$, has been
subtracted to make the minima lie at zero.
The $V^0$ values for $V_\text{discrete}$ and $V_\text{smooth}$
and the the upper bounds of the left and right plots in Figs.\ \ref{figMercator1} -- \ref{figMercator3}
exhibit discrepancies.  They originate from the intrinsic
impossibility of the smooth-tube approximation to correctly account for the actual distribution of the carbon
atoms on the cylinder.  The wider the tube, the more atoms (the higher $N_\text{T}$), and the smaller the
discrepancy: the $(9,9)$ tube ($V^0_\text{smooth} = 70111.5$ K, $V^0_\text{discrete} = -79117.2$ K)
exhibits the largest difference
$V^0_\text{discrete}-V^0_\text{smooth}$ while for the $(21,1)$ tube the values $V^0_\text{discrete}$ and
$V^0_\text{smooth}$ get very close
($V^0_\text{smooth} = -31028.6$ K, $V^0_\text{discrete} = -31033.9$ K).  We point out that any
continuum approach suffers from such a discrepancy, and that it can not be resolved by replacing $\sigma$ in Eq.\
(\ref{smootheq}) by the adjusted density $\tilde{\sigma} = N_\text{T}/(2\pi R_\text{T}2z_\text{max})$, a notion we
illustrate in Appendix \ref{appendixdiscrepancytest}.  Nevertheless, it is not senseless at all to perform a
smooth-tube approach, because conclusions are to be drawn based on the potential energy {\bf variation}: of
interest are the locations of energy minima, corresponding to molecular orientations which are most stable.

\begin{figure}
\resizebox{6cm}{!}{\includegraphics{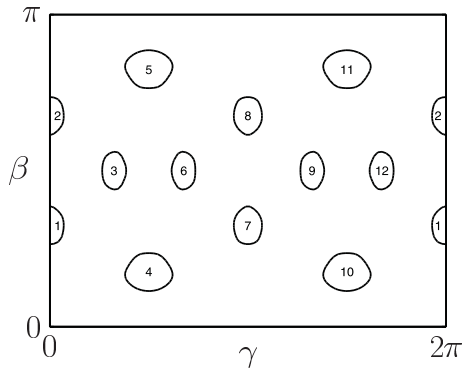}}
\caption{Indication of the locations $(\beta_i,\gamma_i)$, $i=1,\hdots,12$, 
tabulated in Table \ref{anglepairs}. 
For tubes with $R_\text{T}\approx 6.5$ {\AA} (Fig.\ \ref{figMercator1}), 
these angle pairs correspond to the lowest-energy molecular
orientations (Fig.\ \ref{fourtypes}).  The contours are reproductions of the 
$V_\text{smooth} = 200$ K contours of
Fig.\ \ref{figMercator1}(a).
\label{contours_schematically}}
\end{figure}

\begin{figure}
\subfigure{\resizebox{4cm}{!}
{\includegraphics{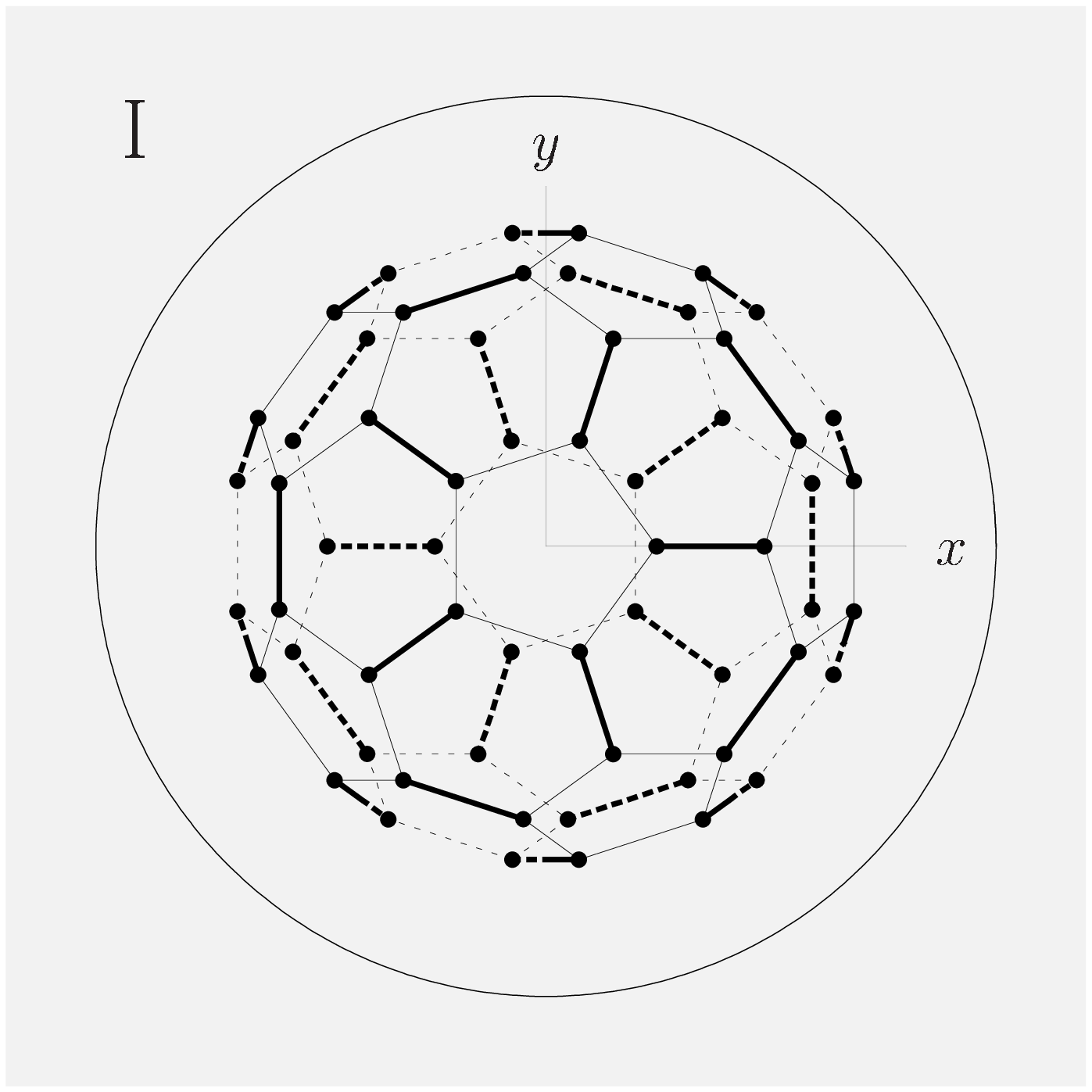}}}
\subfigure{\resizebox{4cm}{!}
{\includegraphics{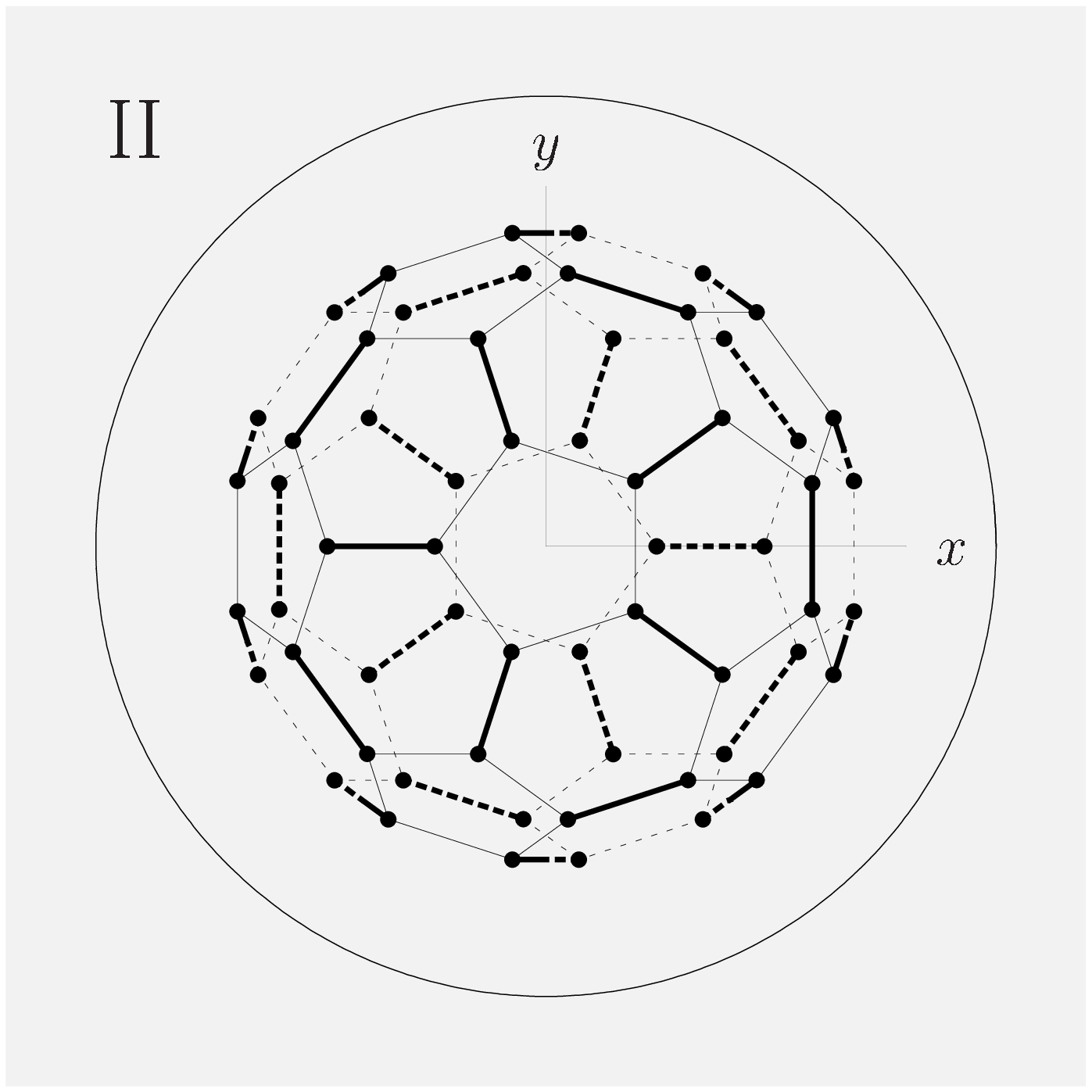}}}
\subfigure{\resizebox{4cm}{!}
{\includegraphics{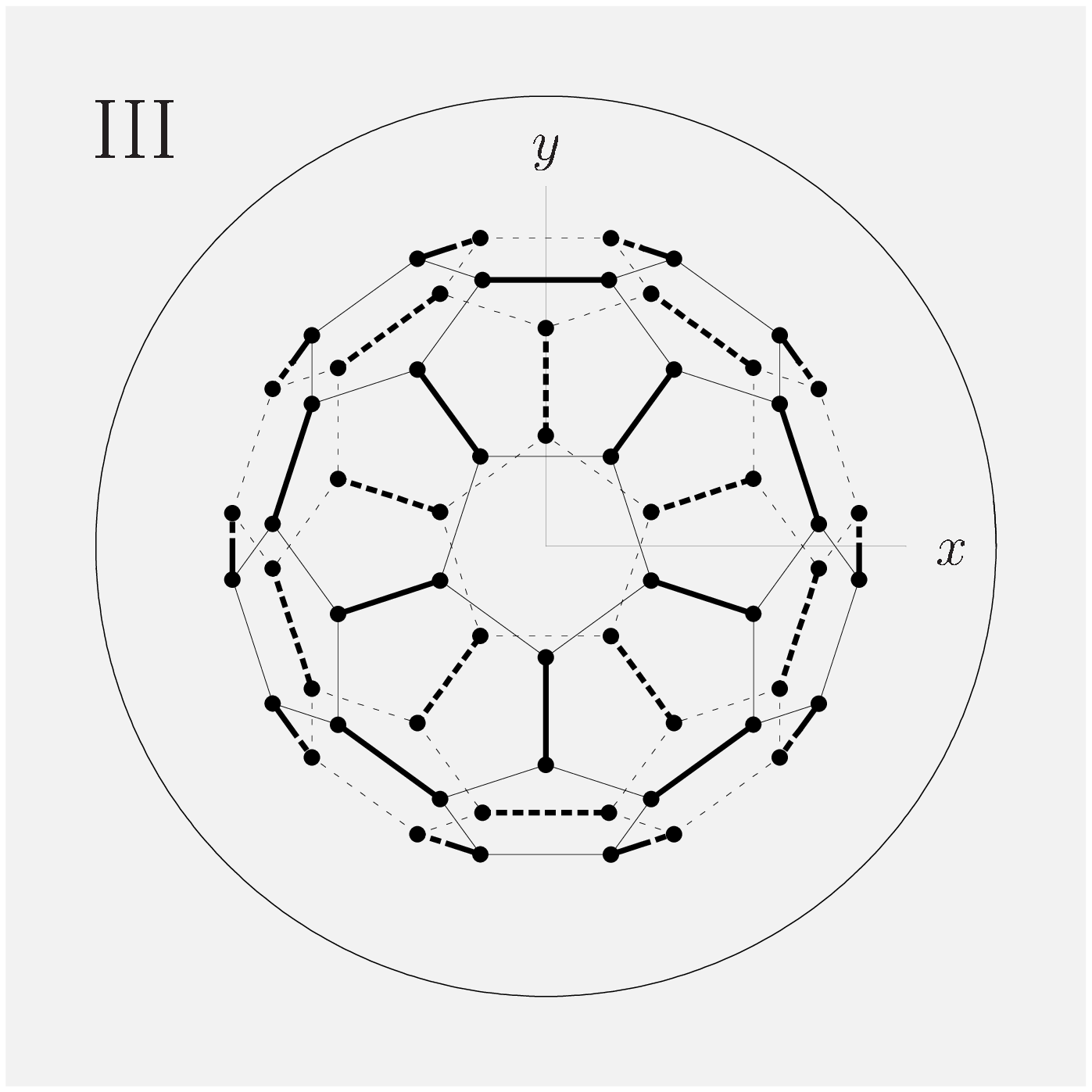}}}
\subfigure{\resizebox{4cm}{!}
{\includegraphics{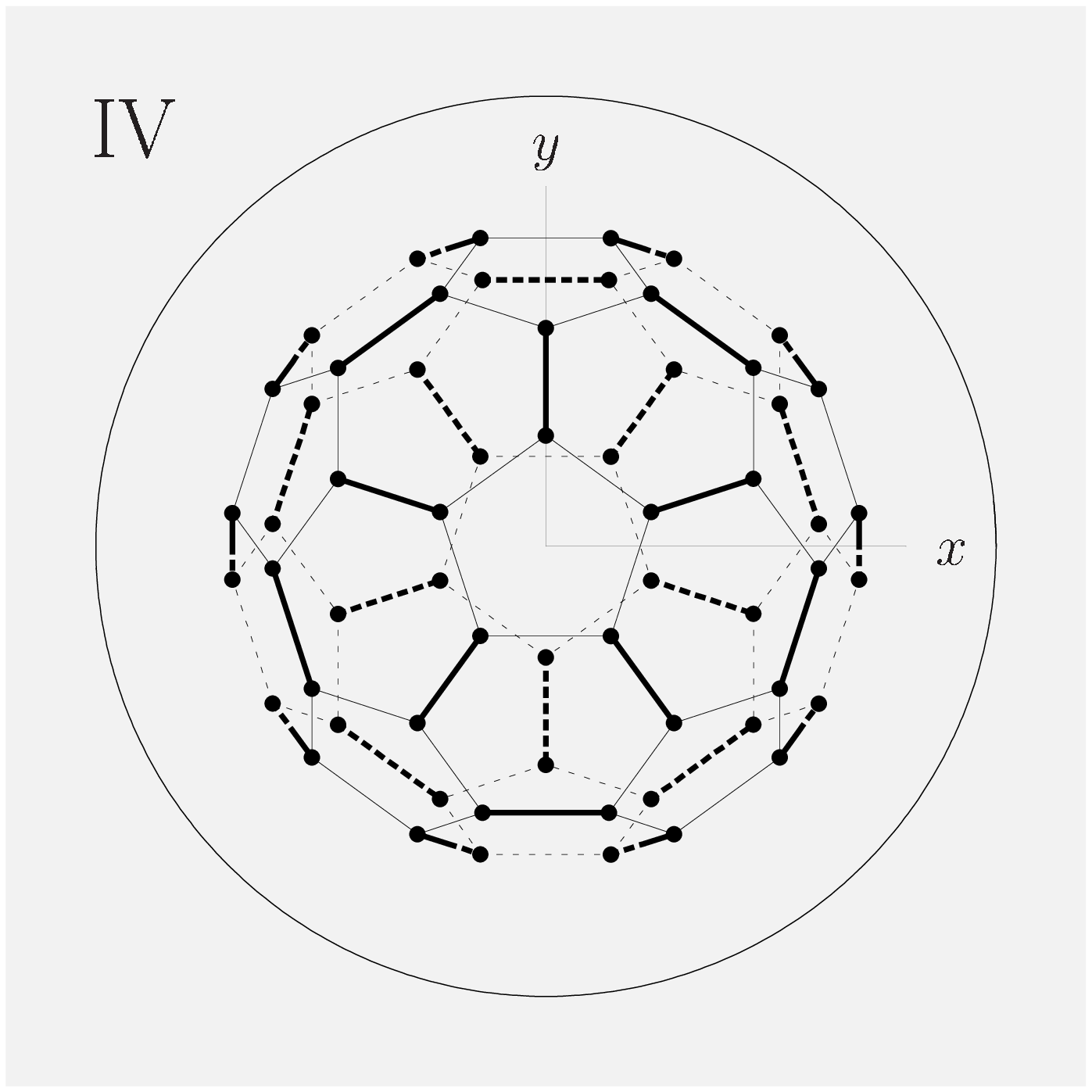}}}
\caption{When applying the 12 Euler transformations ${\mathfrak R}^{-1}(\alpha=0,\beta_i,\gamma_i)$, with
$(\beta_i,\gamma_i)$, $i=1,\hdots,12$, from Table \ref{anglepairs}, only four different molecular orientations
the projections onto the $(x,y)$-plane of which are shown here, are obtained.  Dashed (fragments of) lines have
$z<0$.  The labeling I, II, III and IV correlates with Table \ref{anglepairs}.}
\label{fourtypes}
\end{figure}

The $V_\text{smooth}$ plots in Figs.\ \ref{figMercator1}(a), (b) and (c) are, apart from different energy ranges,
similar.  They exhibit 12 minima (white) and 20 maxima (dark gray).  (The $\gamma$-coordinate, ranging from $0$ to
$2\pi$, is cyclic, i.e.\ the molecular orientations at the $\gamma=0$ ``edge" are repeated at the $\gamma=2\pi$
``edge".  The $\beta$-coordinate is not cyclic: points along the $\beta=0$ and $\beta=\pi$ ``edges" with equal
$\gamma$ refer to distinct configurations.)  The 12 angle pairs $(\beta_i,\gamma_i)$ corresponding to
minimal-energy configurations are tabulated in Table \ref{anglepairs} and indicated schematically in Fig.\
\ref{contours_schematically}.  Each of the 12 minima corresponds to a molecular orientation where two opposing pentagons
of the C$_\text{60}$ molecule are perpendicular to the tube axis [Fig.\ \ref{orientations}(b)]; the Euler transformations
$(\alpha=0,\beta_i,\gamma_i)$, $i=1,\hdots,12$, yield four truly different types of orientations (Fig.\
\ref{fourtypes} and Table \ref{anglepairs}).  The 20 maxima correspond to situations where two facing
hexagons of the C$_\text{60}$ molecule are perpendicular to the $z$-axis [Fig.\ \ref{orientations}(c)].

\begin{table}
\caption{Rotation angles $\beta_i$ and $\gamma_i$, $i=1,\hdots,12$, the 
corresponding Euler transformations
${\mathfrak R}^{-1}(\alpha=0,\beta_i,\gamma_i)$ of which make two facing 
pentagons of the C$_\text{60}$ molecule
lie perpendicular to the $z$-axis.  For small tubes (see Fig.\ 
\ref{figMercator1}),
the minima of $V_\text{smooth}(\beta,\gamma)$ occur at these 12 molecular 
orientations.
Four distinct molecular orientations, labelled I, II, III and IV, are 
obtained,
distinguishable by the orientation of the top ($z>0$) pentagon (Fig.\ 
\ref{fourtypes}).
}
\label{anglepairs}
\begin{ruledtabular}
\begin{tabular}{rrrr}
$i$ & $\beta_i$     & $\gamma_i$  & molecular orientation \\
\hline
$1$  & $\beta_0= \cos^{-1}\frac{2}{\sqrt{10+2\sqrt{5}}}\approx 58^\circ$ & 
$0$              & I   \\
$2$  & $\pi-\beta_0$                                                     & 
$0$              & II  \\
$3$  & $\frac{\pi}{2}$                                                   & 
$\beta_0$        & III \\
$4$  & $\frac{\pi}{2}-\beta_0$                                           & 
$\frac{\pi}{2}$  & II  \\
$5$  & $\frac{\pi}{2}+\beta_0$                                           & 
$\frac{\pi}{2}$  & I   \\
$6$  & $\frac{\pi}{2}$                                                   & 
$\pi-\beta_0$    & IV  \\
$7$  & $\beta_0$                                                         & 
$\pi$            & I   \\
$8$  & $\pi-\beta_0$                                                     & 
$\pi$            & II  \\
$9$  & $\frac{\pi}{2}$                                                   & 
$\pi+\beta_0$    & III \\
$10$ & $\frac{\pi}{2}-\beta_0$                                           & 
$\frac{3\pi}{2}$ & II  \\
$11$ & $\frac{\pi}{2}+\beta_0$                                           & 
$\frac{3\pi}{2}$ & I   \\
$12$ & $\frac{\pi}{2}$                                                   & 
$2\pi-\beta_0$   & IV  \\
\end{tabular}
\end{ruledtabular}
\end{table}

Comparing $V_\text{discrete}$ and $V_\text{smooth}$ of Fig.\ \ref{figMercator1}(a) --- $(16,0)$ tube ---, we see
that the locations of minima and maxima hardly (or even do not) differ.
For the $(9,9)$ armchair tube, Fig.\ \ref{figMercator1}(c), the energy ranges coincide, 
and the minima locations of $V_\text{discrete}$ and $V_\text{smooth}$ are, if not coinciding, almost equal.
Deviations are observed in Fig.\ \ref{figMercator1}(b) for the chiral $(14,4)$ tube: the minima 
locations of $V_\text{discrete}$ clearly
deviate somewhat from those of $V_\text{smooth}$; some
even ``split into two". 
The deviations are small, however: one may write the true $V_\text{discrete}$ 
minima locations as
$(\beta'_i = \beta_i + \Delta\beta_i,\gamma'_i = \gamma_i + \Delta\gamma_i)$.  
We estimate maximal deviation
values at $|\Delta\beta_i|\approx 10^\circ$  and $|\Delta\gamma_i|\approx 
12^\circ$.

For $R_\text{T}\approx 7.5$ {\AA} (Fig.\ \ref{figMercator2}), the $V_\text{discrete}$ plots become extremely
similar to the respective $V_\text{smooth}$ plots.  For all three investigated tubes --- $(19,0)$, $(16,5)$ and
$(11,11)$ --- the locations of minima (and maxima) can be concluded to coincide.  The 30 minima correspond with
two opposing double bonds being perpendicular to the tube axis [Fig.\ \ref{orientations}(a)].  Maximal energy
occurs when two opposing pentagons are perpendicular to the tube axis (the 12 minimal-energy configurations for
$R_\text{T}\approx 6.5$ {\AA}, Fig.\ \ref{figMercator1}).

In Fig.\ \ref{figMercator3}, $R_\text{T}\approx 8.5$ {\AA}, $V_\text{discrete}$ and $V_\text{smooth}$
match completely.  With respect to Fig.\ \ref{figMercator1}, minima and maxima have been flipped: lowest-energy
configurations now feature hexagons perpendicular to the tube axis, pentagons perpendicular to the tube axis yield
the highest energy.

Up to now, $\alpha$ and $\zeta$ have been kept fixed.
To get an idea of the energy variation when $\alpha$ and $\zeta$ are allowed to vary, we have calculated
$V_\text{discrete}(\alpha,\beta = 0,\gamma = 0;\zeta = 0;n,m)$ and
$V_\text{discrete}(\alpha = 0,\beta = 0,\gamma = 0;\zeta;n,m)$, for $0\le \alpha \le \Delta\alpha$ and
$0 \le \zeta \le \Delta\zeta$, respectively.  The tube-dependent rotational and translational periods
$\Delta\alpha$ and $\Delta\zeta$ are given in Table \ref{tubes}.  In Table \ref{alphazetavariation}, we
summarise these
calculations by listing the differences
$\Delta_\alpha V_\text{discrete} = \max\bigl(V_\text{discrete}(\alpha,\beta = 0,\gamma = 0;\zeta = 0;n,m)\bigr)
 - \min\bigl(V_\text{discrete}(\alpha,\beta = 0,\gamma = 0;\zeta = 0;n,m)\bigr)$ and
$\Delta_\zeta V_\text{discrete} = \max\bigl(V_\text{discrete}(\alpha = 0,\beta = 0,\gamma = 0;\zeta;n,m)\bigr)
 - \min\bigl(V_\text{discrete}(\alpha = 0,\beta = 0,\gamma = 0;\zeta;n,m)\bigr)$
conveying the energy variation.  Clearly, the tubes with $R_\text{T}\approx 6.5$ {\AA} display fairly large
energy fluctuations upon varying $\alpha$ and/or $\zeta$, as one might intuitively guess from the jagged contours
in Fig.\ \ref{figMercator1} (left).  For intermediate ($R_\text{T}\approx 7.5$) and large tube radii
($R_\text{T}\approx 8.5$) the energy variations are small.  

\begin{table}
\caption{
Energy variations $\Delta_\alpha V_\text{discrete}$ and
$\Delta_\zeta V_\text{discrete}$ (units K) when varying $\alpha$ or $\zeta$ and keeping other variables fixed.
}
\label{alphazetavariation}
\begin{ruledtabular}
\begin{tabular}{rrr}
$(n,m)$ & $\Delta_\alpha V_\text{discrete}$ (K) & $\Delta_\zeta V_\text{discrete}$ (K) \\
\hline
$(16,0)$  &  $51.3$ & $1079.3$ \\
$(14,4)$  & $643.2$ &  $639.6$ \\
$(9,9)$   & $665.9$ &    $8.3$ \\
\hline
$(19,0)$  &  $<0.1$ &    $3.7$ \\
$(16,5)$  &   $0.4$ &    $0.6$ \\
$(11,11)$ &  $<0.1$ &   $<0.1$ \\
\hline
$(21,0)$  &  $<0.1$ &    $0.1$ \\
$(21,1)$  &  $<0.1$ &    $0.3$ \\
$(12,12)$ &  $<0.1$ &   $<0.1$ \\
\end{tabular}
\end{ruledtabular}
\end{table}

We conclude that the smooth-tube approach works well
for tube radii $R_\text{T}\approx 7.5$ {\AA} and higher, as
seen from the Mercator maps in Figs.\ \ref{figMercator2} and \ref{figMercator3} and the energy variations of Table
\ref{alphazetavariation}.  For small-radius tubes ($R_\text{T}\lesssim 7$ {\AA}), a
systematic investigation addressing the variation of
$V_\text{discrete}$ as a function of $\alpha$, $\beta$, $\gamma$ and $\zeta$ is required.  In the following
section we perform such a study for the $(16,0)$, $(14,4)$ and $(9,9)$ and three more low-radius tubes.

\section{Low-radius tubes: full energy variation}\label{casestudy}
Before proceeding to the full energy variation calculation required for peapods with $R_\text{T}\approx 6.5$
{\AA}, we would like to reflect on actual small values of peapod radii.  To our knowledge, both today's
experimental and theoretical situation do not show unanimity.  Theoretically, different lower limits for a
peapod's radius have been suggested, based on the outcome of the reaction energy $\Delta E$ in the reaction
$(n,m) + \text{C$_\text{60}$}\longrightarrow\text{C$_\text{60}$@$(n,m)$}-\Delta E$: exo- ($\Delta E < 0$) or
endothermic ($\Delta E >0$).  Okada et al.\ \cite{OkadaPRB2003,OkadaPRL2001} concluded from density-functional
theory calculations that
for C$_\text{60}$@$(n,n)$ with $10\le n\le 13$ the reaction is exothermic, and, by extrapolating the results of
$n=8,9$ and $10$, obtained a minimal tube radius of $R_\text{T}^\text{min}\approx 6.4$ {\AA} \cite{OkadaPRL2001}.
Rochefort \cite{Roc}, performing molecular mechanics calculations, set the lower limit at
$R_\text{T}^\text{min}\approx 5.9$ {\AA}.  From the experimental side, while it is still impossible to manufacture
nanotubes --- let alone peapods --- with a given pair of indices $(n,m)$, peapod samples with a narrow radial
dispersion around a mean value $\overline{R}_\text{T}$ and good filling rates (typically, $75\%
$) can be produced at present.  We mention a few (recent) experiments on peapods.  Cambedouzou et al.\
\cite{Cam2005} used a sample with $\overline{R}_\text{T}\approx 6.8$ {\AA}.  Maniwa et al.\ \cite{Maniwa2003}
fitted x-ray diffraction data on C$_\text{60}$@SWCNT peapods to simulations, resulting in a mean radius
$\overline{R}_\text{T}\approx 6.76$ {\AA}.  Kataura et al.\ \cite{KatauraAPA2002} reported measurements on a
sample having a diameter range of $6.25$ {\AA} $\le R_\text{T}\le 7.35$ {\AA}, from which we calculate
$\overline{R}_\text{T}\approx 6.8$ {\AA}.  The electron diffraction studies of Hirahara et al.\ \cite{Hir2001}
were performed on peapod samples with $\overline{R}_\text{T}\approx 7.15$ {\AA} (SWCNTs from a same batch were
used to synthesize not only C$_\text{60}$ -, but also C$_\text{70}$ - and C$_\text{80}$ peapods).  Kataura et al.\
\cite{KatauraSynthMetals2001} reported high-yield fullerene encapsulation, controlling the tubes to be ``larger
than the $(10,10)$ tube", i.e.\ $R_\text{T}^\text{min}\gtrsim 6.86$ {\AA}.  Pfeiffer et al.\ \cite{Pfe} inferred
from Raman spectroscopy their three samples to have $\overline{R}_\text{T}\approx 7$ {\AA},
$\overline{R}_\text{T}\approx 6.52$ {\AA} and $\overline{R}_\text{T}\approx 6.505$ {\AA}.  From all these values
one may conclude that peapods with a radius around $6.5$ {\AA}, our representative value for the ``pentagonal
case" (Figs.\ \ref{figMercator1} and \ref{fourtypes}), although possible, are less abundant than peapods with a
radius around, say, $6.75$ {\AA}.  We have therefore considered a few additional tubes --- $(17,0)$, $(14,5)$ and
$(10,10)$ --- with radii around $6.75$ {\AA}; their characteristics are shown in Table \ref{extratubes}.  These
tubes can be expected to be more realistic representatives of the ``pentagonal" regime instead of the
$R_\text{T}=6.5$ {\AA} tubes of Refs.\ \onlinecite{MicVerNikPRL} and \onlinecite{MicVerNikEPJB} --- we recall that
the transition from the ``pentagonal" to the ``double-bond" lowest-energy orientation occurs around $7$ {\AA}
\cite{MicVerNikEPJB}.
We note that the $(10,10)$ tube is of special interest since
tubes with a radius close to $R_\text{T}(10,10)=6.86$ {\AA} are favorable for C$_\text{60}$ encapsulation, as 
seen in both experiment --- according to
Kataura et al.\ \cite{KatauraAPA2002}, peapod samples tend to have a radial dispersion centered around
$R_\text{T}(10,10)$ --- and theory --- of the C$_\text{60}$@$(n,n)$ series ($n=8,...,13$), the $n=10$
peapod stands  out as the most ``exothermic" (see above) \cite{OkadaPRB2003, OkadaPRL2001}.

\begin{table*}
\caption{Characteristics of additional tubes ($R_\text{T}\approx 6.75$ 
{\AA}).  See captions to Tables \ref{tubes} and \ref{alphazetavariation}.
}
\label{extratubes}
\begin{ruledtabular}
\begin{tabular}{rrrrrrr}
$(n,m)$ & chirality & $R_\text{T}$ ({\AA})& $\Delta\alpha$ & $\Delta\zeta$ ({\AA}) & $\Delta_\alpha
V_\text{discrete}$ (K) & $\Delta_\zeta V_\text{discrete}$ (K) \\
\hline
$(17,0)$  & zig-zag  & $6.7370$ & $2\pi/17$ & $\sqrt{3}a = 4.3128$ & $<0.1$ & $176.4$ \\
$(14,5)$  & chiral   & $6.7603$ & $2\pi$    & $24.5237$            & $36.3$ & $36.5$  \\
$(10,10)$ & armchair & $6.8640$ & $2\pi/10$ & $a = 2.49$           & $23.1$ & $11.6$  \\
\end{tabular}
\end{ruledtabular}
\end{table*}

\begin{figure*}
\resizebox{15cm}{!}{\includegraphics{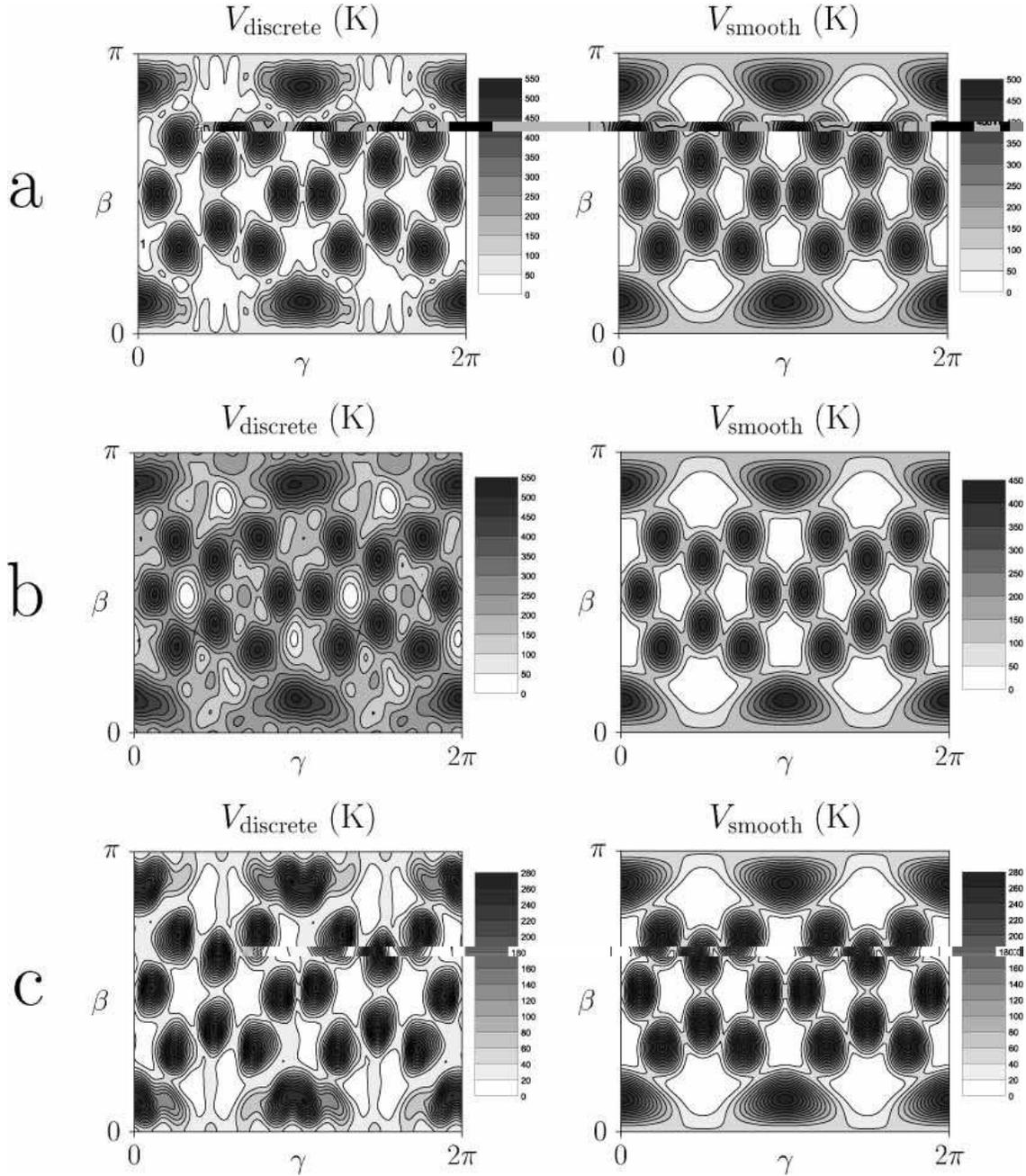}}
\caption{Mercator maps $V_\text{discrete}(\beta,\gamma)$ (left) and 
$V_\text{smooth}(\beta,\gamma)$
(right), units K: (a)
$(n,m) = (17,0)$, (b) $(n,m) = (14,5)$, (c) $(n,m) = (10,10)$.  The minimal 
values have been subtracted.}
\label{figMercatorextra}
\end{figure*}

The Mercator maps $V_\text{smooth}(\beta,\gamma)$ and $V_\text{discrete}(\beta,\gamma)$ of the extra tubes are
shown in Fig.\ \ref{figMercatorextra}; and the energy variations $\Delta_\alpha V_\text{discrete}$ and
$\Delta_\zeta V_\text{discrete}$ are listed in Table \ref{extratubes}.  Again, minimal energies occur around
$(\beta_i,\gamma_i)$, and maxima correspond to hexagons being perpendicular to the tube axis.  The $(17,0)$ tube's
$V_\text{discrete}$ $50$ K contour deviates from its smooth-tube $50$ K contour [Fig.\ \ref{figMercatorextra}(a)],
but there is some over-all agreement.  The $(10,10)$ tube's $V_\text{discrete}$ plot [Fig.\
\ref{figMercatorextra}(c), left] features ``split" minima as seen for the $(14,4)$ tube [Fig.\
\ref{figMercator1}(b)].  The $(14,5)$ tube's $V_\text{discrete}$ plot does not coincide nicely with its
$V_\text{smooth}$ plot, but interestingly, the two locations $(\beta_i,\gamma_i)$, $i=3$ and $i=9$, type III
``pentagonal" orientations, correspond very well to their smooth-approximation counterparts.  Since all other ten
minima locations are related to the $i=3$ or the $i=9$ location by a molecular rotation over
$\alpha=\frac{\pi}{4}$ or $\alpha=\frac{\pi}{2}$ about the $z$-axis --- see Fig.\ \ref{fourtypes} ---, any of the
$(\beta_i,\gamma_i)$ points can be made a minimal configuration by changing $\alpha$.  The main conclusion is that
the minimal-energy orientation will always feature facing pentagons perpendicular to the $z$-axis.  The same can
be said of any of the tubes of Figs.\ \ref{figMercator1} and \ref{figMercatorextra} --- tubes with
$R_\text{T}\lesssim 7$ {\AA} --- excepting the $(14,4)$ tube.

We now turn to the $\alpha$- and $\zeta$-dependencies of $V_\text{discrete}$.
It is sufficient to consider the intervals
$0\le\alpha<\Delta\alpha\equiv\alpha_\text{max}$ and $0\le\zeta<\Delta\zeta\equiv\zeta_\text{max}$.
We divide the interval $[0,\alpha_\text{max}[\times[0,\zeta_\text{max}[$ into a
$10\times 10$ grid $(\alpha_{i=1,\hdots,10},\zeta_{j=1,\hdots,10})$, $\alpha_i=(i-1)\frac{\alpha_\text{max}}{10}$,
$\zeta_j=(j-1)\frac{\zeta_\text{max}}{10}$, and construct a double 
Fourier series --- for notational simplicity we drop the indices $(n,m)$ ---,
\begin{multline}
   V_\text{discrete}(\alpha,\beta,\gamma;\zeta) \\ =
      \sum_{p=0}^{\infty}\sum_{q=0}^{\infty}\Biggl\{
A_{pq}(\beta,\gamma)\cos\biggl(p\frac{\alpha}{\alpha_\text{max}}2\pi\biggr)
\cos\biggl(q\frac{\zeta}{\zeta_\text{max}}2\pi\biggr)
      \displaybreak[0] \\
      +
B_{pq}(\beta,\gamma)\sin\biggl(p\frac{\alpha}{\alpha_\text{max}}2\pi\biggr)
\cos\biggl(q\frac{\zeta}{\zeta_\text{max}}2\pi\biggr) \\
      +
C_{pq}(\beta,\gamma)\cos\biggl(p\frac{\alpha}{\alpha_\text{max}}2\pi\biggr)
\sin\biggl(q\frac{\zeta}{\zeta_\text{max}}2\pi\biggr)
      \displaybreak[0] \\
      +
D_{pq}(\beta,\gamma)\sin\biggl(p\frac{\alpha}{\alpha_\text{max}}2\pi\biggr)
\sin\biggl(q\frac{\zeta}{\zeta_\text{max}}2\pi\biggr)
      \Biggr\}, \label{series}
\end{multline}
by numerically calculating the Fourier coefficients
\begin{multline}
   A_{pq}(\beta,\gamma) \\ = N_{pq}
      \int_0^{\alpha_\text{max}}d\alpha\,
      \int_0^{\zeta_\text{max}}d\zeta\,
      V_\text{discrete}(\alpha,\beta,\gamma;\zeta) \\
      \times \cos\biggl(p\frac{\alpha}{\alpha_\text{max}}2\pi\biggr)
      \cos\biggl(q\frac{\zeta}{\zeta_\text{max}}2\pi\biggr), \\
\hdots \nonumber
\end{multline}
via the trapezium-rule approximations
\begin{multline}
   A_{pq}(\beta,\gamma) \\ \approx 
N_{pq}\sum_{i=1}^{10}\sum_{j=1}^{10}\frac{\alpha_\text{max}}{10}
      \frac{\zeta_\text{max}}{10}
      V_\text{discrete}(\alpha_i,\beta,\gamma;\zeta_j) \\
      \times \cos\biggl(p\frac{\alpha_i}{\alpha_\text{max}}2\pi\biggr)
      \cos\biggl(q\frac{\zeta_j}{\zeta_\text{max}}2\pi\biggr), \\
\hdots \nonumber
\end{multline}
The coefficients $B_{pq}$, $C_{pq}$ and $D_{pq}$ are obtained by replacing $\cos()\cos()$ by $\sin()\cos()$,
$\cos()\sin()$ and $\sin()\sin()$, respectively.
The prefactors read
\begin{align}
   N_{00} & = \frac{1}{\alpha_\text{max}\zeta_\text{max}}, \nonumber \\
   N_{0s} & = N_{r0} = \frac{2}{\alpha_\text{max}\zeta_\text{max}}, \nonumber \\
   N_{rs} & = \frac{4}{\alpha_\text{max}\zeta_\text{max}}, \label{Fprefactors}
\end{align}
where $r$ and $s\ne0$.  In series (\ref{series}), some terms may vanish because of symmetry reasons.

\begin{figure*}
\includegraphics{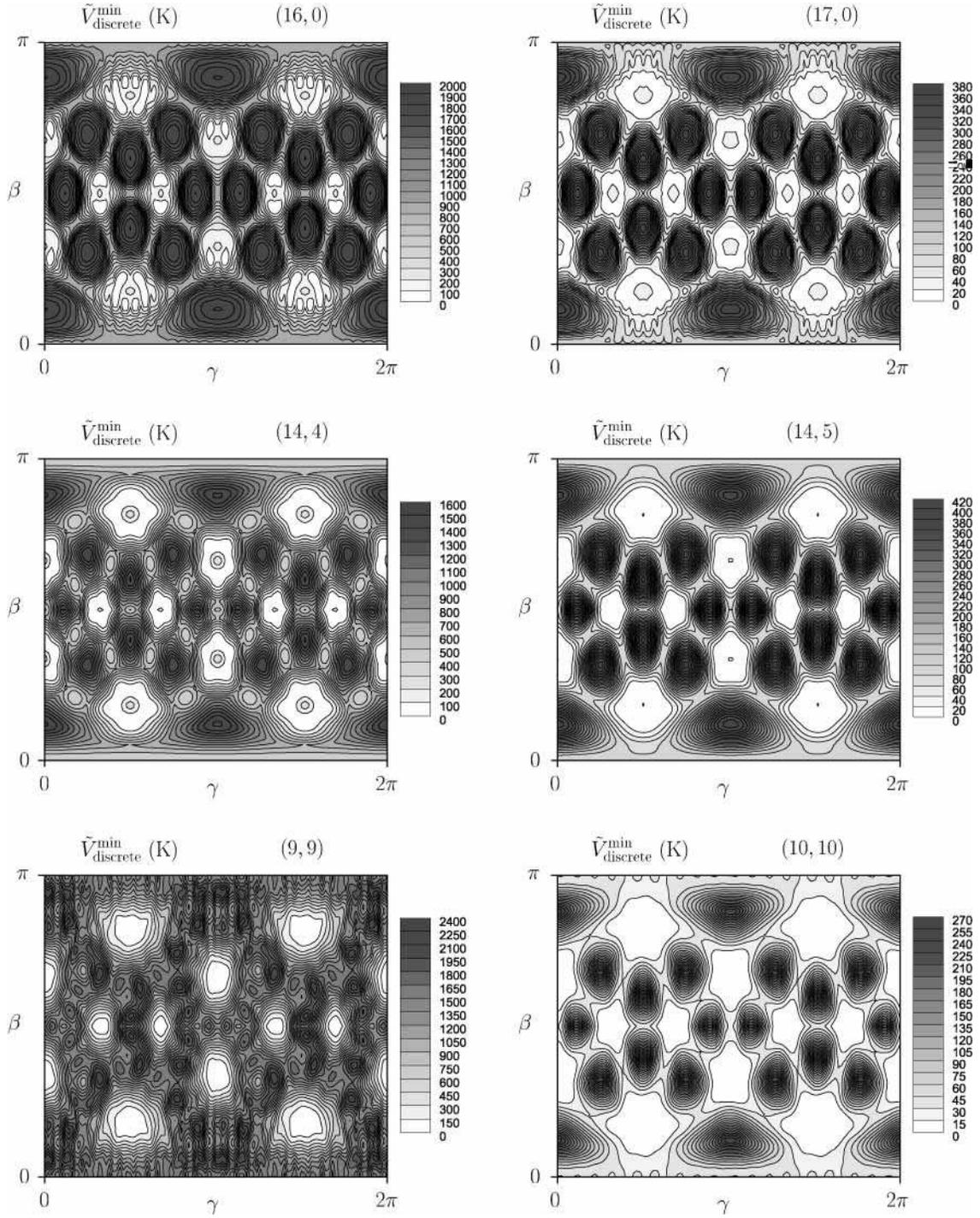}
\caption{Mercator maps $\tilde{V}_\text{discrete}(\beta,\gamma)$, units K, for $(16,0)$, $(14,4)$, $(9,9)$,
$(17,0)$, $(14,5)$ and  $(10,10)$ peapods.  The minimal values have been subtracted.
}
\label{fullvariation}
\end{figure*}

The Fourier coefficients $A_{pq}(\beta,\gamma)$, $B_{pq}(\beta,\gamma)$, $C_{pq}(\beta,\gamma)$,
$D_{pq}(\beta,\gamma)$ can be interpreted as Mercator maps.  The magnitude of
the coefficients decreases for increasing indices $p$ and $q$; we
approximate $V_\text{discrete}(\alpha,\beta,\gamma;\zeta)$ by
\begin{multline}
   \tilde{V}_\text{discrete}(\alpha,\beta,\gamma;\zeta) =
      \sum_{p=0}^4\sum_{q=0}^4\Biggl\{
A_{pq}(\beta,\gamma) \\
\times\cos\biggl(p\frac{\alpha}{\alpha_\text{max}}2\pi\biggr)
\cos\biggl(q\frac{\zeta}{\zeta_\text{max}}2\pi\biggr)
      + \hdots
      \Biggr\}.
\end{multline}
For given $\beta$ and $\gamma$, we scan the $\alpha$- and $\zeta$-intervals and define
\begin{multline}
   \tilde{V}_\text{discrete}^\text{min}(\beta,\gamma) \equiv
   \min\bigl\{\tilde{V}_\text{discrete}(\alpha,\beta,\gamma;\zeta); \\
   0\le \alpha\le \alpha_\text{max}0\le
   \zeta\le \zeta_\text{max}\bigr\}.
\end{multline}
The quantity $\tilde{V}_\text{discrete}^\text{min}(\beta,\gamma)$,
gives the lowest 
attainable energy $\tilde{V}_\text{discrete}$ when varying $\alpha$ and $\zeta$.
In Fig.\ \ref{fullvariation}, it has been plotted for every of the six tubes investigated.
The plots again exhibit icosahedral symmetry as in the previous Mercator maps.  The main observation here is
that for all tubes, except the $(9,9)$ tube [Fig.\ \ref{fullvariation}(c), left], the absolute minima lie not
precisely at the 12 $(\beta_i,\gamma_i)$ locations but somewhat away from them --- the same effect observed for
the $(\alpha=0,\zeta=0)$ V$_\text{discrete}$ plots in Figs.\ \ref{figMercator1} and \ref{figMercatorextra}.  As
before, we can write the actual minimum locations as $(\beta'_i = \beta_i+\Delta\beta_i,\gamma'_i = \gamma_i+\Delta\gamma_i)$.  For the
$R_\text{T}\approx 6.5$ {\AA} tubes (Fig.\ \ref{fullvariation}, left), excepting the $(9,9)$ tube, the
$(\beta_i,\gamma_i)$ locations (orientations) have energies $\sim 300$ K higher than the minimal energies.  The
$(9,9)$ tube's minima are really close to --- if not, coinciding with --- the $(\beta_i,\gamma_i)$ orientations. 
For the $R_\text{T}\approx 6.75$ {\AA} tubes (Fig.\ \ref{fullvariation}, right), excepting the $(10,10)$ tube,
the $(\beta_i,\gamma_i)$ orientations have energies $\sim 20$ K higher than the minimal energies.  The $(10,10)$
tube's absolute minima lie also off the $(\beta_i,\gamma_i)$ ``pentagon" orientations (not visible on the plot),
and have
energies $\sim 12$ K higher than the lowest energies.
We must therefore conclude that chirality-dependent effects manifest themselves here.
However, for $R_\text{T}\approx 6.75$ {\AA} tubes, probably the smallest peapod tubes as discussed above,
the effects can be said to be minor.  As an approximation, one may consider the smooth-tube approach.
We recall that for higher tube radii, the smooth-tube approximation is excellent (when the C$_\text{60}$
molecules lie on the tube axis).

To conclude this section, we come back to the 4 types of ``pentagonal" orientations depicted
in Fig.\ \ref{fourtypes}.  We recall that they arise from the Euler transformations $\mathfrak{R}^{-1}(\alpha =
0,\beta_i,\gamma_i)$ with the angle pairs $(\beta_i,\gamma_i)$, $i=1,\hdots,12$, of Table \ref{anglepairs}.
Clearly, all 12 $V_\text{smooth}(\beta_i,\gamma_i)$ values are identical because of the cylindrical symmetry.  A
priori, $V_\text{discrete}(\beta_i,\gamma_i)$ can be different for each of the four types I, II, III and IV.
Interestingly, depending on the tube's symmetry, some orientations sometimes are equivalent.  This is illustrated
in Table \ref{equivalent}, and can be understood by being aware of certain symmetry elements.
Orientations I and II are related via a rotation over $\pi$ about the $z$-axis, and likewise for orientations III
and IV.  Orientations III and I are related by a rotation over $\pi/2$ about the $z$-axis.  The presence of a
twofold symmetry axis for the $(n,m)$ tube therefore implies equivalence of orientations I and II and of
orientations III and IV, while a fourfold axis implies the equivalence of all orientations.  The $(16,0)$ tube is
an example of the latter, a $(14,4)$ tube provides an example of the former.  The $(16,5)$ tube has no pure
rotational symmetry axis and exhibits therefore distinguishable $V_\text{discrete}$ values.
We note that any occurring energy differences as a result of the
discussed inequivalences are small, however, and not seen on any of the Mercator maps since the contour
values lie not sufficiently close to each
other.  Also note that these observations are generally true 
for the values
$V_\text{discrete}(\beta_i,\gamma_i)$ --- they do not need to be to minima, 
e.g.\ the $(14,4)$ tube [Fig.\
\ref{figMercator1}(b)].
The equivalence relationships between the twelve $(\beta_i,\gamma_i)$ orientations is nicely seen in the
the $(14,5)$ tube's $V_\text{discrete}$ plot [Fig.\ \ref{figMercatorextra}(b)]: all 4 types I ($\left\{1,5,7,11\right\}$), 
II ($\left\{2,4,8,10\right\}$), III ($\left\{3,9\right\}$) and IV ($\left\{6,12\right\}$) ``behave" differently.
When making the observations concerning the  equality/inequality of
$V_\text{smooth}(\beta_i,\gamma_i)$ values summarized schematically in Table \ref{equivalent}, we have made sure
that sufficient numerical accuracy has been achieved.  One needs accurate enough atomic coordinates of the
C$_\text{60}$ molecule, so that upon explicitly applying coordinate transforms 
${\mathfrak R}^{-1}(\alpha,\beta,\gamma)$ corresponding to symmetry elements of the C$_\text{60}$ molecule (i.e.,
of the icosahedral group $I_h$), the same set of coordinates is obtained up to the desired accuracy.  We use coordinates with
12 significant digits.

\begin{table*}
\caption{
Schematical presentation of $V_\text{discrete}(\beta_i,\gamma_i)$ values for selected tubes, indicating equalities
and inequalities due to the presence/absence of certain symmetries.  The angles $\beta_i$ and $\gamma_i$ are
tabulated in Table \ref{fourtypes}, the corresponding molecular orientations depicted in Fig.\ \ref{anglepairs}.
}
\label{equivalent}
\begin{ruledtabular}
\begin{tabular}{rrrr}
$i$ (type) & \multicolumn{3}{c}{$V_\text{discrete}(\beta_i,\gamma_i)$} \\
    & $(n,m) = (16,0)$ & $(n,m) = (14,4)$ & $(n,m) = (16,5)$  \\
\hline
$1,5,7,11$ (I)  & $V^{(16,0)}_1$ & $V^{(14,4)}_1$ & $V^{(16,5)}_1$ \\
$2,4,8,10$ (II) & $V^{(16,0)}_1$ & $V^{(14,4)}_1$ & $V^{(16,5)}_2$ \\
$3,9$ (III)     & $V^{(16,0)}_1$ & $V^{(14,4)}_3$ & $V^{(16,5)}_3$ \\
$6,12$ (IV)     & $V^{(16,0)}_1$ & $V^{(14,4)}_3$ & $V^{(16,5)}_6$
\end{tabular}
\end{ruledtabular}
\end{table*}

\section{Discussion and conclusions}\label{discussion}
We have presented a systematic comparison of the potential energy of a C$_\text{60}$ molecule --- treated as an
icosahedral cluster of ICs --- encapsulated centrally in a SWCNT, when approximating the tube as a uniform
cylinder and when taking the true carbon atomic network into account.  The former approach results in only two
variables (the Euler angles $\beta$ and $\gamma$), while the latter requires in addition a third Euler angle
$\alpha$ and the molecule's $z$-coordinate $\zeta$ denoting its position along the tube axis.  The
$(\beta,\gamma)$-dependence can be conveniently plot as a Mercator map.  Fixing $(\alpha,\zeta)$ at $(0,0)$ then
allows a first visual comparison of the ``smooth" and ``discrete" Mercator maps $V_\text{smooth}(\beta,\gamma)$
and $V_\text{discrete}(\beta,\gamma)$.  From these preliminary comparisons one can see that the larger the tube,
the better the smooth-tube approximation.  Indeed, the $V_\text{smooth}$ and $V_\text{discrete}$ Mercator maps for
tubes with radii $R_\text{T}\gtrsim 7.5$ {\AA} (Figs.\ \ref{figMercator2} and Fig.\ \ref{figMercator3}) are as
good as identical.  For smaller tubes, the effect of the tube structure comes into play and deviations
between the smooth and discrete Mercator maps are visible (Figs.\ \ref{figMercator1} and Fig.\
\ref{figMercatorextra}).  While the $(\alpha,\zeta)$-dependence of $V_\text{discrete}$ can be argued to be
negligible for the larger tubes ($R_\text{T}\gtrsim 7$ {\AA}) because of the similarity of
$V_\text{smooth}(\beta,\gamma)$ and $V_\text{discrete}(\beta,\gamma)$, a full investigation of the
variables $\alpha$ and $\zeta$ is in order for smaller tubes.  We have presented a detailed study for selected
zig-zag, chiral and armchair tubes with radii around $R_\text{T}\approx 6.5$ {\AA} and $6.75$ {\AA}, including the
$(10,10)$ tube, nowadays considered the ideal peapod tube \cite{KatauraAPA2002,OkadaPRB2003,OkadaPRL2001}.

A double Fourier series captures the $(\alpha,\zeta)$-dependence in a manageable way.
Scanning $V_\text{discrete}(\alpha,\beta,\gamma;\zeta;n,m)$ for its lowest attainable values when varying
$\alpha$ and $\zeta$ yields Mercator maps similar to $V_\text{smooth}(\beta,\gamma;n,m)$.  We see that the actual
energy minima do not correspond to the 12 ``pentagonal" $(\beta_i,\gamma_i)$ orientations but that they lie
slightly away from them (except
for the $(9,9)$ tube, where the minimal-energy molecular configurations are really close to
the ``pentagonal" orientations).
Such $(\beta,\gamma)$ orientations correspond to ``tilted" molecules, where an axis connecting the midpoints of two
opposing pentagons does not coincide with the tube's long axis ($z$-axis) but makes a cone with a small opening
angle if one would perform the $\alpha$ Euler rotation.  Hence, we conclude that for these smaller tubes (again
excepting the $(9,9)$ tube), the chirality of the tube does play a role.  However, as seen from the small
deviations occurring in the Mercator plots, the effect is not strong, especially for the $R_\text{T}\approx 6.75$
{\AA} tubes (Fig.\ \ref{fullvariation}, right).  Therefore, one may consider the smooth-tube
approach if one wants to capture radius-dependent properties.
In this respect, one should be aware of the present-day experimental situation:
precise knowledge of the components (i.e., occurring chiralities) in a peapod sample is absent --- only a
determination of the tube
radius distribution seems feasible up to now.  The main conclusion of Refs.\
\onlinecite{MicVerNikPRL} and \cite{MicVerNikEPJB} stands:
three regimes can be distinguished
($R_\text{T}\lesssim 7$ {\AA},
``pentagonal" orientation, $7$ {\AA} $\lesssim R_\text{T}\lesssim 7.9$ {\AA}, ``double-bond" orientation and
$R_\text{T}\gtrsim 7.9$ {\AA}, ``hexagonal" orientation).  The ``pentagonal" orientations have to be restated as
``tilted pentagonal" orientations, though.
Generally, our findings are in accordance with those of Troche et
al.\ \cite{Troche} who concluded that the chirality of the SWCNT encapsulating the C$_\text{60}$ molecules has
only a minor effect.

Note that transversal motion of the C$_\text{60}$ molecule (off-axis displacements) is not discussed here since
our sole purpose was a direct comparison of the smooth-tube and the discrete-tube approaches for a centrally
located C$_\text{60}$ molecule. 
As expected intuitively and demonstrated in Refs.\ \onlinecite{MicVerNikPRL} and \onlinecite{MicVerNikEPJB}, from
a certain radius ($R_\text{T}\approx 7$ {\AA}) on, an off-center position is energetically more favorable.  This,
however, involves energy differences much larger than those seen upon varying $\beta$, $\gamma$, $\alpha$ and
$\zeta$.


The smooth-tube approximation's
requiring only the two Euler angles $\beta$ and $\gamma$ allows for the use of symmetry-adapted rotator functions. 
For details we refer to Refs.\ \onlinecite{MicVerNikPRL} and \onlinecite{MicVerNikEPJB}.  The advantage lies in
the possibility of writing $V_\text{smooth}(\beta,\gamma;R_\text{T})$ as an expansion into functions
${\mathfrak U}_l(\beta,\gamma)$ taking the icosahedral molecular
symmetry of the C$_\text{60}$ molecule and the cylindrical
symmetry of its site into account,
\begin{align}
   V_\text{smooth}(\beta,\gamma;R_\text{T}) = \sum_{l=0}^{\infty}w_l(R_\text{T}){\mathfrak U}_l(\beta,\gamma), \label{expansionintoSARFs}
\end{align}
where the coefficients $w_l$ relate to the icosahedral symmetry and 
carry the details of the pair interaction potential.  The molecular symmetry
implies the first non-vanishing terms to be those with $l=0$, $l=6$, $l=10$ and $l=12$, and a restriction to only
these four leading terms already approximates $V_\text{smooth}(\beta,\gamma;R_\text{T})$ extremely well.  Apart
from providing
mathematical/physical clarity, expansion (\ref{expansionintoSARFs}) greatly reduces the calculation time.  While,
for a $100\times 100$ $(\beta,\gamma)$ grid, an implementation of Eq.\ (\ref{smootheq}) takes hours,
the calculation of $V_\text{smooth}(\beta,\gamma;R_\text{T})$ via Eq.\ (\ref{expansionintoSARFs}) is a matter of
minutes on the same machine.

It is interesting to note that double-walled carbon nanotubes (DWCNTs) allow the encapsulation of C$_\text{60}$
molecules in inner spaces smaller than observed for SWCNTs: Khlobystov et al.\ \cite{Khl} reported the insertion
of C$_\text{60}$ molecules in DWCNTs with internal radii as small as $5.5$ {\AA}.  Having different minimal
internal radii of SWCNTs and DWCNTs for filling with C$_\text{60}$ molecules is attributed to the difference in
how a C$_\text{60}$ molecule interacts with a SWCNT and a DWCNT \cite{Khl}.  We have carried out calculations for
C$_\text{60}$@DWCNT by treating the field on the C$_\text{60}$ molecule as a superposition of the two fields from
the tubes with different radii.  Although we find that the presence of a second (outer) tube decreases the energy
for encapsulation when taking a tube radius difference equalling the interlayer distance of graphite ($3.35$
{\AA}), the effect is rather small and not sufficient to explain the large reduction in inner tube radius.


We believe that the general conclusion --- the smooth-tube approximation being justified for intermediate and
large tube radii ($R_\text{T}\gtrsim 7$ {\AA}) and possibly acceptable for smaller
tube-radii --- reached here is relevant for other peapod systems.  For example, (C$_\text{70}$)$_N$@SWCNT peapods
feature different orientations of the encapsulated C$_\text{70}$ molecules for different tube radii
\cite{Maniwa2003,Hir2001}, the so-called ``lying" (for smaller tube radii) and ``standing" (larger tube radii)
orientations.  A smooth-tube approach would make a
good start for investigating these specific orientations.


\begin{acknowledgments}
We thank A.V. Nikolaev for providing high-precision atomic C$_\text{60}$ coordinates.
B.V.\ is a research assistant of the Fonds voor Wetenschappelijk Onderzoek -- Vlaanderen.
\end{acknowledgments}

\appendix
\section{Discrepancy test}\label{appendixdiscrepancytest}
We consider a single carbon atom in a short $(16,0)$ tube fragment defined by $|z_\tau|\le a/\sqrt{3}\approx 1.42$
{\AA}, leaving only three ``rings" of $16$ carbon atoms each.  The tube atoms have coordinates
$\vec{r}_\tau =(x_\tau,y_\tau,z_\tau)$, $\tau = 1,\hdots,48$.  The single atom, put at the center of the fragment
which we define to be the origin of the employed cartesian coordinates system, has a ``discrete" energy
\begin{align}
   V_\text{discrete} = \sum_{\tau = 1}^{48} v^\text{a}(r_\tau), \label{A1}
\end{align}
approximated by the ``smooth" energy
\begin{align}
   V_\text{smooth} = \sigma R_\text{T}\int_0^{2\pi}d\Phi\int_{Z_\text{min}}^{Z_\text{max}}dZv^\text{a}(\vec{\rho}).
   \label{A2}
\end{align}
In both equations, the pair potential $v^\text{a}(r)$ of Sec.\ \ref{nanotubefield} is understood.
In Eq.\ (\ref{A2}), $\Phi$ and $Z$ are defined via $x = R_\text{T}\cos\Phi$, $y = R_\text{T}\sin\Phi$ and $z = Z$.
The integration boundaries $Z_\text{min}$ and $Z_\text{max}$ are not well-defined.  Indeed, there is a range of
both lower and upper boundaries corresponding to a ``smooth" tube fragment containing only the three ``rings" with
$z_\tau = -a/\sqrt{3}$, $z_\tau = 0$ and $z_\tau = a/(2\sqrt{3})$: $-\sqrt{3}a/2<Z_\text{min}\le-a/\sqrt{3}$ and
$a/(2\sqrt{3})\le Z_\text{max}<\sqrt{3}a/2$.  In Table \ref{discrepancytable} we present $V_\text{smooth}$ values
calculated for a few of these $[Z_\text{min},Z_\text{max}]$ intervals.  For each case, both the tube surface
density $\sigma = 4/(\sqrt{3}a^2)$ and the adjusted tube density
$\tilde{\sigma} = 48/\bigl(2\pi R_\text{T}(Z_\text{max}-Z_\text{min})\bigr)$ has been considered.  The ``discrete"
value, obtained via Eq.\ (\ref{A1}), reads $V_\text{discrete} = -318.1$ K, and is best reproduced by the ``smooth"
value if the interval $[Z_\text{min},Z_\text{max}]=[-\frac{5a}{4\sqrt{3}},\frac{a}{\sqrt{3}}]$ is chosen.  The
tube fragment edges then lie precisely in the middle of two neighboring ``rings" of atoms; the surface densities
$\sigma$ and $\tilde{\sigma}$ then happen to coincide.  The values of table \ref{discrepancytable} suggest that
some choice(s) of intervals may yield the $V_\text{discrete}$ value, but the point we want to make here is that
making use of the adjusted surface density $\tilde{\sigma}$ does not ``convert" the $V_\text{smooth}$ to the
$V_\text{discrete}$ value.
We remark that doing the $V_\text{smooth}$ calculations described in the paper with $\tilde{\sigma}$ instead of
$\sigma$ turned out to yield only very small differences.

\begin{table}
\caption{
$V_\text{smooth}$ values, units K, for various tube fragments and surface densities $\sigma$ and $\tilde{\sigma}$.
}
\label{discrepancytable}
\begin{ruledtabular}
\begin{tabular}{rrr}
$[Z_\text{min},Z_\text{max}]$ & $\sigma$ or $\tilde{\sigma}$ & $V_\text{smooth}$ (K) \\
\hline
$[-\frac{\sqrt{3}a}{2},\frac{\sqrt{3}a}{2}]$ & $\sigma$         & $-405.0$ \\
$[-\frac{\sqrt{3}a}{2},\frac{\sqrt{3}a}{2}]$ & $\tilde{\sigma}$ & $-303.7$ \\
$[-\frac{5a}{4\sqrt{3}},\frac{a}{\sqrt{3}}]$ & $\sigma$         & $-317.0$ \\
$[-\frac{5a}{4\sqrt{3}},\frac{a}{\sqrt{3}}]$ & $\tilde{\sigma}$ & $-317.0$ \\
$[-\frac{a}{\sqrt{3}},\frac{a}{2\sqrt{3}}]$  & $\sigma$         & $-217.1$ \\
$[-\frac{a}{\sqrt{3}},\frac{a}{2\sqrt{3}}]$  & $\tilde{\sigma}$ & $-325.7$ \\
\end{tabular}
\end{ruledtabular}
\end{table}



\begin{thebibliography}{99}

\bibitem{Iij}
S. Iijima, Nature (London) {\bf 354}, 56 (1991).

\bibitem{Ebb}
T.W. Ebbesen and P.M. Ajayan, Nature (London) {\bf 358}, 220 (1992).

\bibitem{Saibook}
R. Saito, G. Dresselhaus, and M.S. Dresselhaus, {\it Physical Properties of Carbon Nanotubes} (Imperial College
Press, London, 1998).

\bibitem{Har}
P.J.F. Harris, {\it Carbon Nanotubes and Related Structures} (Cambridge University Press, Cambridge, 1999).

\bibitem{Slo}
J. Sloan, A.I. Kirkland, J.L. Hutchison and M.L.H. Green, Chem. Commun., 1319 (2002).

\bibitem{Mon}
M. Monthioux, Carbon {\bf 40}, 1809 (2002).

\bibitem{Smi}
B.W. Smith, M. Monthioux, and D.E. Luzzi, Nature (London) {\bf 396}, 323 (1998).

\bibitem{Hor}
D.J. Hornbaker, S.J. Kahng, S. Misra, B.W. Smith, A.T. Johnson, E.J. Mele, D.E. Luzzi, and A. Yazdani, Science
{\bf 295}, 829 (2002).

\bibitem{Smi2}
B.W. Smith, M. Monthioux, and D.E. Luzzi, Chem. Phys. Lett. {\bf 315}, 31 (1999).

\bibitem{Bur}
B. Burteaux, A. Claye, B.W. Smith, M. Monthioux, D.E. Luzzi, and J.E. Fischer, Chem. Phys. Lett. {\bf 310}, 21
(1999).

\bibitem{Pickett}
G.T. Pickett, M. Gross, and H. Okuyama, Phys. Rev. Lett. {\bf 85}, 3652 (2000).

\bibitem{Hod1}
M. Hodak and L.A. Girifalco, Phys. Rev. B {\bf 67}, 075419 (2003).

\bibitem{Khl}
A. Khlobystov, D.A. Britz, A. Ardavan, and G.A.D. Briggs, Phys. Rev. Lett. {\bf 92}, 245507 (2004).

\bibitem{Troche}
K.S. Troche, V.R. Coluci, S.F. Braga, D.D. Chinellato, F. Sato, S.B. Legoas, R. Rurali, and D.S. Galvao,
Nano Lett. {\bf 5}, 349 (2005).

\bibitem{MicVerNikPRL}
K.H. Michel, B. Verberck, and A.V. Nikolaev, Phys. Rev. Lett. {\bf 95}, 185506 (2005).

\bibitem{MicVerNikEPJB}
K.H. Michel, B. Verberck, and A.V. Nikolaev, Eur. Phys. J. B {\bf 48}, 113 (2005).

\bibitem{Mickelson}
W. Mickelson, S. Aloni, W.Q. Han, J. Cummings, and A. Zettl, Science {\bf 300}, 467 (2003).

\bibitem{BraCrack}
C.J. Bradley and A.P. Cracknell, {\it The Mathematical Theory of Symmetry in Solids} (Clarendon, Oxford, 1972).

\bibitem{DavNature}
W.I.F. David, R.M. Ibberson, J.C. Matthewman, K. Prassides, T.J.S. Dennis, J.P. Hare, H.W. Kroto, R. Taylor, and
D.R.M. Walton, Nature (London) {\bf 353}, 147 (1991).

\bibitem{LamZeitschrift}
D. Lamoen and K.H. Michel, Z. Phys. B {\bf 92}, 323 (1993);
J.R.D. Copley and K.H. Michel, J. Phys.: Condens. Matter {\bf 5}, 4353 (1993).

\bibitem{124}
P. Launois, S. Ravy, and R. Moret, Phys. Rev. B {\bf 55}, 2651 (1997).

\bibitem{130}
K.H. Michel and J.R.D. Copley, Z. Phys. B {\bf 103}, 369 (1997).

\bibitem{Ham}
N. Hamada, S. Sawada, and A. Oshiyama, Phys. Rev. Lett. {\bf 68}, 1579 (1992);
D.H. Robertson, D.W. Brenner, and J.W. Mintmire, Phys. Rev. B {\bf 45}, 12592 (1992).


\bibitem{Mercator}
Gerardus Mercator (1512 - 1594), Flemish cartographer, inventor of the cylindrical projection.  The angles
$\beta$ and $\gamma$ play the role of geographical lattitude and longitude, respectively.

\bibitem{OkadaPRB2003}
S. Okada, M. Otani, and A. Oshiyama, Phys. Rev. B {\bf 67}, 205411 (2003).

\bibitem{OkadaPRL2001}
S. Okada, S. Saito, and A. Oshiyama, Phys. Rev. Lett. {\bf 86}, 3835 (2001).

\bibitem{Roc}
A. Rochefort, Phys. Rev. B {\bf 67}, 115401 (2003).

\bibitem{Cam2005}
J. Cambedouzou, V. Pichot, S. Rols, P. Launois, P. Petit, R. Klement, H. Kataura, and R. Almairac, Eur. Phys. J. B
{\bf 42}, 31 (2004); J. Cambedouzou, S. Rols, R. Almairac, J.-L. Sauvajol, H. Kataura, and H. Schober, Phys. Rev. B {\bf 71},
041403(R) (2005).

\bibitem{Maniwa2003}
Y. Maniwa, H. Kataura, M. Abe, A. Fujiwara, R. Fujiwara, H. Kira, H. Tou, S. Suzuki, Y. Achiba, E. Nishibori,
M. Takata, M. Sakata, and H. Suematsu, J. Phys. Soc. Jap. {\bf 72}, 45 (2003).

\bibitem{KatauraAPA2002}
H. Kataura, Y. Maniwa, M. Abe, A. Fujiwara, T. Kodama, K. Kikuchi, H. Imahori, Y. Misaki, S. Suzuki, and
Y. Achiba, Appl. Phys. A: Mater. Sci. Process. {\bf 74}, 349 (2002).

\bibitem{Hir2001}
K. Hirahara, S. Bandow, K. Suenaga, H. Kato, T. Okazaki, H. Shinohara, and S. Iijima, Phys. Rev. B {\bf 64},
115420 (2001).

\bibitem{KatauraSynthMetals2001}
H. Kataura, Y. Maniwa, T. Kodama, K. Kikuchi, K. Hirahara, K. Suenaga, S. Iijima, S. Susuki, Y. Achiba, and
W. Kr\"{a}tschmer, Synth. Met. {\bf 121}, 1195 (2001).

\bibitem{Pfe}
R. Pfeiffer, H. Kuzmany, T. Pichler, H. Kataura, Y. Achiba, M. Melle-Franco, and F. Zerbetto, Phys. Rev. B
{\bf 69}, 035404 (2004).

\end{thebibliography}
\end{document}